\documentclass[aps,amsmath,amssymb,prb,twocolumn,superscriptaddress,showpacs]{revtex4}
\usepackage{bm}
\usepackage{epsfig}
\usepackage[usenames]{color}
\usepackage{graphicx}

\newcommand{\nn}{\nonumber}
\newcommand{\paper}{paper }

\newcommand{\sgn}{\operatorname{sgn}}

\begin{document}

\title{Thermoelectric and Seebeck coefficients of granular metals}

\author{Andreas~Glatz}
\affiliation{Materials Science Division, Argonne National Laboratory, Argonne, Illinois 60439, USA}

\author{I.~S.~Beloborodov}
\affiliation{Department of Physics and Astronomy, California State University, Northridge, California 91330, USA}

\date{\today}

\begin{abstract}
In this work we present a detailed study and derivation of the thermopower and thermoelectric coefficient of nano-granular metals at large tunneling conductance between the grains, $g_{T}\gg 1$.
An important criterion for the performance of a thermoelectric device is the thermodynamic figure of merit which
is derived using the kinetic coefficients of granular metals.
All results are valid at intermediate temperatures, $E_c\gg T/g_{T} > \delta$, where $\delta$ is the mean energy
level spacing for a single grain and $E_c$ its charging energy.
We show that the electron-electron interaction leads to an increase of the thermopower with decreasing grain size and discuss our results in the light of future generation thermoelectric materials for low temperature applications.
The behavior of the figure of merit depending on system parameters like grain size, tunneling conductance, and temperature is presented.
\end{abstract}

\maketitle

\section{introduction}

Thermoelectric materials with high efficiency is a major research area in condensed matter physics and materials science for several decades now.
Due to recent advances in nano-fabrication, these materials promise next generation devices for conversion of thermal energy to electrical energy and vice versa.
A measure for the performance or efficiency of a thermoelectric material is the dimensionless {\it figure of merit}, usually denoted as $ZT$, where $T$ is the temperature. It depends on the thermopower or Seebeck coefficient, and the electric and thermal conductivities~\cite{Rowe,Abrikosov,Mahanbook,Mahan, Bell}. However, the Wiedemann-Franz law has defeated much progress in increasing the performance of bulk materials, since it directly relates electric and thermal conductivity whereas the figure of merit is proportional to the quotient of both.
The Seebeck coefficient of a material measures the magnitude of an induced thermoelectric voltage in response to a temperature difference across that material. If the temperature difference $\Delta T$ between the two ends of a material is small, then the thermopower, $S$, of the material is defined as $S = - \Delta V/\Delta T$, where $\Delta V$ is the voltage difference across a sample.

In general, thermoelectric devices are used as converters for either electrical power into heating/cooling ({\it Peltier effect}) or sources of different temperature into electricity ({\it Seebeck effect}). These devices are usually much simpler, especially without moving parts, than conventional devices, e.g. two-phase compressors for cooling, and therefore more reliable. However, for both effects the materials need to have good electrical conductivity to minimize ohmic heating and at the same time to be bad thermal conductors to avoid thermal equilibration of the temperature gradient. Therefore, the aim is to create materials which optimize these parameters together with the thermopower. Currently thermoelectric devices based on p- and n-type-doped semiconductor junctions archive only about 12\% of the maximal theoretical efficiency (as compared to 60\% in conventional cooling systems)~\cite{Bell}.

To be competitive compared with conventional refrigerators, one must develop thermoelectric materials with large $ZT$. The highest figure of merit for bulk thermoelectric materials is about $1$, but in order to match the efficiency of mechanical systems, $ZT\sim 9$ is needed~\cite{Bell}.
However, for $ZT\gtrsim 2$ thermoelectic applications become economically competitive~\cite{disalvo99,sales02,snyder08,tritt06}.
Although it is possible in principle to develop homogeneous materials with that large figure of merit, there are no concrete devices on the horizon.
Especially promising for further improvement in efficiency are {\it inhomogeneous/granular} thermoelectric materials~\cite{Majumdar} in which one can directly control the system parameters. In Ref.~[\onlinecite{venka01}] a figure of merit at $300K$ of $2.4$ for a layered nanoscale structure and later in Ref.~[\onlinecite{harman02}] a $ZT$ of $3.2$ at about $600K$ for a bulk material with nanoscale inclusions were reported.

Overall, recent years have seen a remarkable progress in the design of granular conductors with controllable structure parameters.  Granules can be capped with organic (ligands) or inorganic molecules which connect and regulate the coupling between them.  Altering the size and shape of granules one can regulate quantum confinement effects. In particular, tuning microscopic parameters one can vary the granular materials from
being relatively good metals to pronounced insulators as a function of the strength of electron tunneling couplings between conducting grains~\cite{Collier97,Murray,Tran}.  This makes granular conductors a perfect exemplary system for studying thermoelectric and related phenomena.

All these experimental achievements and technological prospects call for a comprehensive theory able to provide quantitative description of not only the electric but also thermoelectric properties of granular conductors, which can in future serve as basis for a clever design of devices for a new generation of nano-thermoelectrics.

Most theoretical progress so far was archived by numerical solution of phenomenological models~\cite{Linke,Broido}. However, no analytical results obtained from a microscopic model for coupled nanodot/grain systems is available up to now. Thus, the fundamental question that remains open is how thermoelectric coefficient and thermopower behave in nanogranular thermoelectric materials.  Here, we make a step towards answering this question for granular metals at intermediate temperatures by generalizing our approach~\cite{Beloborodov07} recently developed for the description of electric~\cite{Beloborodov03} and heat transport~\cite{Beloborodov05}.  In particular, we will answer the question to what extend quantum and confinement effects in nanostructures are important in changing $ZT$.

In this \paper we investigate the thermopower $S$, thermoelectric coefficient $\eta$, and the figure of merit $ZT$ of granular samples focusing on the case of large tunneling conductance between the grains, $g_{T}\gg 1$.
Without Coulomb interaction the granular system would be a good metal in this limit and our task is to include charging effects in the theory. We furthermore restrict our considerations to the case of intermediate temperatures
\begin{equation}
E_c \gg T/g_T> \delta,
\end{equation}
where $\delta$ is the mean level spacing of a single grain and $E_c$ the charging energy. The left inequality means that the temperature is not high enough such that Coulomb effects are pronounced. It also allows us to perform all calculations up to logarithmic accuracy. The right inequality allows us to consider the electronic motion as coherent within the grains, however this coherence does not extend to scales larger than the size of a single grain~\cite{Beloborodov07}.
In Ref.~[\onlinecite{glatz+prb09}] we presented a few major results which we extend here and describe the derivations in much more detail.

The paper is organized as follows: In Sec.~\ref{sec.results} we summarize our main results and discuss their range of applicabilities, in Sec.~\ref{sec.model} we introduce the model, and in Sec.~\ref{sec.thermcoefficient} we outline the derivation of thermoelectric coefficient of granular metals in and without the presence of interaction which is the main result of this paper. In the following section we discuss the behavior of  the Seebeck coefficient and figure of merit (Sec.~\ref{sec.SZ}) as a function of sample parameters. Finally, in Sec.~\ref{sec.discussions} we discuss our findings and present possible further applications of our method.
Important details of our calculations are presented in several comprehensive Appendixes: in Appendix~\ref{app.hom} we calculate the thermoelectric coefficient of homogeneous disorder metals in the absence of interaction. In Appendix~\ref{app.cur} we derive the heat and electric currents of granular metals, and in Appendixes~\ref{app.eta0} and \ref{app.eta1} we calculate the thermoelectric coefficient of granular metals without and in the presence of interaction, respectively.

\section{Results and Summary}
\label{sec.results}

In this section we summarize our results and discuss their range of applicabilities.
The main results of our work are as follows: (i)  We derive the expression for the thermoelectric coefficient $\eta$ of granular metals that includes corrections due to Coulomb interaction at temperatures $T> g_T \delta$, where $\delta$ is the mean level spacing of a single grain
\begin{subequations}
\begin{equation}\label{eq.eta}
\eta =\eta ^{(0)}\left( 1-\frac{1}{4 g_{T} d} \, \ln \frac{g_T E_c}{T} \right).
\end{equation}
Here
\begin{equation}
\eta^{(0)} = - (\pi^2/3)e g_T a^{2-d} (T/\varepsilon _{F}),
\end{equation}
\end{subequations}
is the thermoelectric coefficient
of granular materials in the absence of electron-electron interaction with $e$ being the electron charge, $a$ the size of a single grain, $d= 2,3$ the dimensionality of a sample, $\varepsilon_F$ being the Fermi energy, and $E_c=e^2/a$ is the charging energy.

The condition for the temperature range of our theory ensures that the argument of the logarithm in Eq.~(\ref{eq.eta}) is much larger than $1$, such that all numerical prefactors under the logarithm can be neglected. Furthermore, it also defines a critical lower limit for the grain size when the charging energy $E_c$ becomes of order of the mean energy level spacing $\delta$.

At the temperatures under consideration, the electron motion is coherent
within the grains, but coherence does not extend to scales
larger than the size $a$ of a single grain~\cite{Beloborodov07}. Under these
conditions, the electric conductivity $\sigma$ and the electric thermal conductivity $\kappa$ are given by the expressions~\cite{Beloborodov03,Efetov02,Beloborodov05}
\begin{subequations}
\begin{eqnarray}
\frac{\sigma}{\sigma^{(0)}} &=& 1 - \ln(g_TE_c/T)/(2\pi d g_T),
\label{eq.sigma}\\
\frac{\kappa}{\kappa^{(0)}} &=&  1 - \frac{\ln [g_T E_c/T]}{2\pi d g_T} + \frac{1}{2\pi^2 g_T}\left\{
\begin{array}{lr}
 3\gamma,
\hspace{0.8cm} d=3 &  \\
\ln \frac{g_{T} E_c }{T}, \hspace{0.1cm} d=2
\end{array}
\right. . \label{eq.kappa}
\end{eqnarray}
where
\begin{equation}
\sigma^{(0)} = 2 e^{2}g_{T}a^{2-d} \hspace{0.3cm} {\rm and} \hspace{0.3cm} \kappa^{(0)} = L_0 \sigma^{(0)} T,
\end{equation}
\end{subequations}
are the electric (including spin) and thermal conductivities of granular metals in the absence of Coulomb interaction with $L_0 = \pi^2/3 e^2$ being the Lorentz number. We mention that at temperature $T > g_T \delta$ the correction to the thermoelectric coefficient, Eq.~(\ref{eq.eta}), has a $T \ln T$
dependence in both $d=2,3$ dimensions which is similar to the result for the electric conductivity, Eq.~(\ref{eq.sigma}), having a $\ln T$ dependence in all dimensions as well.

(ii) Using the above results, we obtain the expression for thermopower $S$ of granular metals
\begin{subequations}
\begin{equation}\label{eq.S}
S = S^{(0)}\left(1 - \frac{\pi -2}{4 \pi g_T d} \ln \frac{g_T E_c}{T} \right),
\end{equation}
where
\begin{equation}
S^{(0)} = - (\pi^2/6) (1/e) (T/\varepsilon_F),
\end{equation}
\end{subequations}
is the thermopower of granular metals in the absence of Coulomb interaction.

(iii) Finally, we find the figure of merit to be:
\begin{subequations}
\begin{equation}\label{eq.ZT}
\frac{Z}{Z^{(0)}} = 1 - \frac{\pi - 2}{2 \pi g_T d}  \ln \frac{g_T E_c}{T} -
\frac{1}{2\pi^2 g_T}\left\{
\begin{array}{lr}
 3 \gamma,
\hspace{0.8cm} d=3  \\
\ln \frac{g_{T} E_c }{T}, \hspace{0.1cm} d=2
\end{array}
\right.,
\end{equation}
where
\begin{equation}\label{eq.Z0}
Z^{(0)} T = (\pi^2/12) (T/\varepsilon_F)^2,
\end{equation}
\end{subequations}
is the bare figure of merit of granular materials and $\gamma \approx 0.355$ is a numerical coefficient.
In Sec.~\ref{sec.SZ} we present plots of $Z$ in dependence of various sample parameters.  We find, that the influence of granularity is most effective for small grain sizes and the presence of Coulomb interaction decreases the figure of merit.

At this point we remark that all results are obtained in the absence of phonons which become relevant only at higher temperatures. At the end of this \paper we will briefly discuss their influence.

\section{Model}
\label{sec.model}

We start our considerations with the introduction of our model.
We consider a $d-$dimensional array of metallic grains with Coulomb interaction between electrons.
The motion of electrons inside the grains is diffusive, i.e., the electron's mean free path $\ell$ is smaller than the grain size $a$, and they tunnel from grain to grain. We assume that the sample would be a good metal in the absence of Coulomb interaction.
However, we also assume that the tunneling conductance $g_{T}$ is still smaller than the grain conductance $g_{0}$,
meaning that the granular structure is pronounced and the
resistivity is controlled by tunneling between grains.

Each grain is characterized by two energy scales: (i) the mean
energy level spacing $\delta$, and (ii)
the charging energy $E_c = e^2/a$ (for a typical grain size of $a \approx
10$nm $E_c$ is of the order of $2000$K) and we assume that the condition
$\delta \ll E_c$ is fulfilled.

The system of coupled metallic grains is described by the Hamiltonian $\hat{H} =
\sum _i \hat{H}_i$, where the sum is taken over all grains in the system and
\begin{equation}
\label{Hamiltonian}
\hat{H}_i =  \sum_{k}\xi_k \hat{a}^{\dagger}_{ik} \hat{a}_{ik} + \sum_{j \ne i}\,\frac{e^{2}\hat{n}_{i}\hat{n}_{j}}{2C_{ij}}\,
 + \sum_{j,p,q} (t_{ij}^{pq} \hat{a}^{\dagger}_{ip} \hat{a}_{jq} +\text{c.c.}) .
\end{equation}
The first term on the right hand side (r.~h.~s.)
of Eq.~(\ref{Hamiltonian}) describes the $i$-th isolated disordered grain, $\hat{a}^{\dagger}_{i,k} (\hat{a}_{i,k})$ are the creation
(annihilation) operators for an electron in the state $k$ and $\xi_k = k^2/2m - \mu$
with $\mu$ being the chemical potential.

The second term describes the charging energy, $C_{ij}$ is the capacitance matrix and $\hat{n}_{i} = \sum_k \hat{a}^{\dagger}_{ik} \hat{a}_{ik}$ is the number operator for
electrons in the $i$-th grain. The Coulomb interaction is long ranged and its off-diagonal components cannot be neglected. Note that, since metallic grains have an infinite dielectric constant, the effective dielectric constant of the whole sample can be considerably larger than the dielectric constant of its insulating component. Thus the effective single-grain charging energy can be much less than the electrostatic energy of a single grain in vacuum.

The last term in Eq.~(\ref{Hamiltonian}) is the tunneling part of the Hamiltonian
where $t_{ij}$ are the tunnel matrix elements between grains $i$ and $j$ which we consider to be random Gaussian variables defined
by their correlators:
\begin{equation}
\label{def_t^2}
 \langle \; t^{*p_1q_2}_{ij} \, t^{p_1^\prime q_2^\prime}_{ij}
\;\rangle = t^2_{ij} \; \delta_{p_1 p_1^\prime} \; \delta_{q_2
q_2^\prime },
\end{equation}
$t^2_{ij} = t^2_0 = const$. The dimensionless tunneling conductance is
related to the average matrix elements as
\begin{equation}
g_T = 2\pi  (t^2_0/\delta^2).
\end{equation}
The conductance $g_T$ is defined per one spin component, such that,
for example,  the high temperature (Drude) conductivity of a
periodic granular sample is $\sigma^{(0)} = 2 e^2 g_T a^{2-d}$.

\section{Thermoelectric coefficient}
\label{sec.thermcoefficient}

Having introduced our model in the previous section, we come now to the main methodical part of our work, the derivation of the thermoelectric coefficient $\eta$.
In general the three kinetic coefficients are: the electric conductivity $\sigma$, the thermoelectric coefficient $\eta$, and the thermal conductivity $\kappa$.
They are related to the Matsubara response functions $L^{(\alpha\beta)}$ with $\alpha,\beta\in\{e,h\}$~\cite{Mahan,Mahanbook,Jonson} as
\begin{subequations}
\begin{eqnarray}
{\bf j}^{(e)}& =& - (L^{(ee)}/(e^2 T))\, \nabla (e V) - (L^{(eh)}/(e T^2))\,  \nabla T, \nonumber \\
{\bf j}^{(h)} &=& - (L^{(eh)}/(e T))\, \nabla (e V) - (L^{(hh)}/T^2) \, \nabla T.
\label{responce}
\end{eqnarray}
Here ${\bf j}^{(e)}$ (${\bf j}^{(h)}$) is the electric (thermal) current and $V$ the electrostatic potential.
From Eq.~(\ref{responce}) one finds that
\begin{eqnarray}
\label{00}
\sigma = L^{(ee)}/T , \hspace{0.8cm} \eta = L^{(eh)}/T^2, \\
S = -\Delta V / \Delta T = L^{(eh)}/(T L^{(ee)}), \nonumber
\end{eqnarray}
\end{subequations}
where the response functions are given by Kubo formulas
\begin{equation}
L^{(\alpha\beta)} =  -\left.\frac{\imath T \, \partial }{a^{d} \partial \Omega }\right|_{\Omega \rightarrow 0}  \left[ \int\limits_0^{1/T}d\tau \, e^{\imath\Omega_m \tau} \langle T_{\tau}{\bf j}^{(\alpha)}(\tau) {\bf j}^{(\beta)}(0)\rangle \right]_{\Omega_m \rightarrow - \imath\Omega + \delta}\,,
\end{equation}
with $T_{\tau}$ being the time ordering operator for the currents with respect to the imaginary time $\tau$.
Thus, to calculate the thermoelectric coefficient
$\eta$ and thermopower $S$ of granular metals one has to know
the explicit form of the electric ${\bf j}^{(e)}$ and thermal ${\bf j}^{(h)}$ currents.

\begin{figure}[t]
\includegraphics[width=0.6\columnwidth]{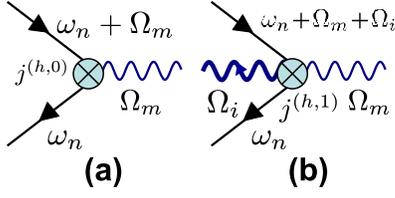}
\caption{(color online) Vertices corresponding to the thermal current operator, Eqs.~(\ref{thermalvertex}):
vertex (a) corresponds to $\widehat{\jmath }_{ij}^{(h,0)}$ and (b)
to $\widehat{\jmath }_{ij}^{(h,1)}$. The solid lines denote the propagator of electrons, the thick wavy line describes Coulomb interaction, the tunneling vertices are described by the circles, $\omega_n = \pi T(2n+1)$ and $\Omega_m =2\pi m T$ are Fermionic and Bosonic Matsubara frequencies respectively ($n,m\in\mathbb{Z}$).}
\label{fig.jver}
\end{figure}

The electric current ${\bf j}_i^{(e)}$ through grain $i$ is defined as
\begin{subequations}
\begin{equation}
{\bf j}^{(e)}_i = \sum_j \hat{\jmath }_{ij}^{(e)} = e\, d\hat{n}_{i}/dt = \imath e[\hat{n}_{i}, \hat{H]}.
\end{equation}
Straightforward calculations (see Appendix~\ref{app.cur}) lead to
\begin{equation}
\hat{\jmath }_{ij}^{(e)}=\imath e\underset{k,q}{\sum }
( t_{ij}^{kq} \hat{a}_{ik}^{\dagger } \hat{a}_{jq}-t_{ji}^{qk} \hat{a}_{jq}^{\dagger } \hat{a}_{ik}).
\end{equation}
\end{subequations}
For granular metals the thermal current operator
\begin{subequations}
\begin{equation}
{\bf j}_i^{(h)} = \sum_j \hat{\jmath }_{ij}^{(h)},
\end{equation}
can be obtained as follows. The energy content of each grain changes as a function
of time, such that $d\hat{H}_i/dt = i [\hat{H}_i,\hat{H}]$.
Energy conservation requires that this energy flows to other grains in the system, $d\hat{H}_{i}/dt \equiv \sum_j \hat{\jmath }_{ij}^{(h)}$.
Calculating the
commutator $[\hat{H}_i,\hat{H}]$, we obtain (for details see Appendix~\ref{app.cur})
\begin{eqnarray}
\label{thermalvertex}
\hat{\jmath }_{ij}^{(h)} &=& \hat{\jmath }_{ij}^{(h,0)} + \hat{\jmath }_{ij}^{(h,1)}, \\
\hat{\jmath }_{ij}^{(h,0)}&=&\imath \underset{k,q}{\sum }
\frac{\xi _{k}+\xi _{q}}{2}\left[ t_{ij}^{kq} \hat{a}_{ik}^{\dagger }
\hat{a}_{jq}-t_{ji}^{qk} \hat{a}_{jq}^{\dagger } \hat{a}_{ik}
\right],\label{12c} \\
\hat{\jmath }_{ij}^{(h,1)}&=&-\frac{e}{4}\underset{m}{
\sum } \left[ \frac{\{ \hat{n}_{i};\hat{\jmath }
_{jm}^{(e)} \} _{+}}{C_{im}}-\frac{ \{ \hat{n}_{j};\hat{\jmath }
_{im}^{(e)} \} _{+}}{C_{jm}} \right],
\end{eqnarray}
\end{subequations}
where $\{ \hat{A};\hat{B} \} _{+}$ denotes the anti-commutator.
The contribution $\widehat{\jmath }_{ij}^{(h,0)}$ is the heat current in the absence of electron-electron interaction, while the second term $\hat{\jmath }_{ij}^{(h,1)}$ appears due to Coulomb interaction. Equation~(\ref{thermalvertex}) implies that the thermal current operator must be associated with two different vertices in diagram representation, Fig.~\ref{fig.jver}.
We remark that Eq.~(\ref{Hamiltonian}) suggests also a finite contribution to ${\bf j}_i^{(h)}$ proportional to $t^2$ -- which indeed exists. However, it vanishes when summed over the sample (for details see Appendix~\ref{app.cur}).

For large tunneling conductance, the Matsubara thermal
current - electric current correlator can be analyzed perturbatively in
$1/g_T$, using the diagrammatic technique discussed in
Ref.~[\onlinecite{Beloborodov07}] that we briefly outline
below. The self-energy of the averaged single electron Green's
function has two contributions: The first one corresponds
to scattering by impurities inside a single grain while the second
is due to the process of scattering between the grains. The former
results only in a small renormalization of the relaxation time which depends in general on the electron energy $\omega$ as

\begin{equation}
\label{tau}
\tau _{\omega }^{-1} = \tau _{0}^{-1}\left[ 1+(d/2-1)\omega /\varepsilon _{F}\right],
\end{equation}
which is a result of the renormalization of the density of states at the Fermi surface [see Eqs.~(\ref{01}) and (\ref{02})].

In the following we outline the calculation of thermoelectric coefficient $\eta$ in the non-interacting case and its correction due to Coulomb interaction. A detailed derivation of both can be found in Appendices~\ref{app.eta0} and~\ref{app.eta1}, respectively.

\subsection{non-interacting case}

\begin{figure}[t]
\includegraphics[width=\columnwidth]{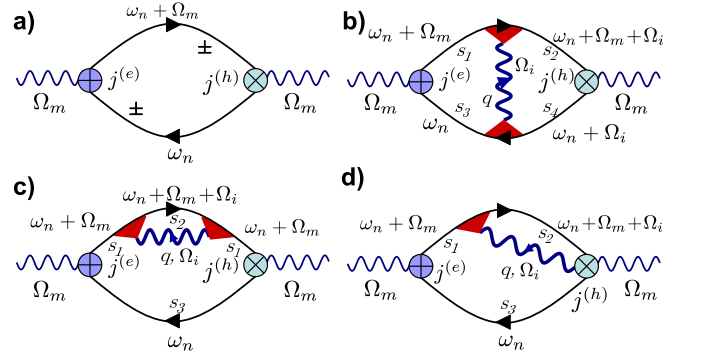}
\caption{(color online) Diagrams describing the thermoelectric
coefficient of granular metals at temperatures $T > g_T\delta$:
diagram (a) corresponds to $\protect\eta_{0}$
in Eq.~(\ref{eq.eta}). Diagrams (b)-(d) describe first order corrections
to the thermoelectric coefficient of granular metals $\eta^{(1)}$ in Eq.~(\ref{eta1sum1})
due to electron-electron interaction. The solid lines denote the propagator of electrons,
the wavy lines describe effective screened electron-electron
propagator, and the (red) triangles describe the elastic
interaction of electrons with impurities. The tunneling vertices are described by the circles.
The sum of the diagrams (b)-(d) results in the thermoelectric coefficient
correction $\eta^{(1)}$ given in Eq.~(\ref{eta1}).}
\label{fig.dia}
\end{figure}

First, we consider the thermoelectric coefficient $\eta^{(0)}$ of granular metals in the absence of interaction. The expression for the thermoelectric coefficient in linear response theory is
\begin{equation}
\label{eq.eta0grain}
\eta ^{(0)}=\imath \left. \frac{\partial }{a^{d}T\partial \Omega }%
\right\vert _{\Omega =0}Q^{(0)}.
\end{equation}
Here $a$ is the grain size and $Q^{(0)}$ the correlator of the heat current, $\overrightarrow{\jmath }^{(h,0)}$ (see Fig.~\ref{fig.jver}a), and electric current, $\overrightarrow{\jmath }^{(e)}$, shown in Fig.~\ref{fig.dia}a.
Notice, that there is an important difference between calculations of the thermoelectric coefficient $\eta$ and thermopower $S$ and the calculations of the electric $\sigma$ and thermal $\kappa$ conductivities in Eqs.~(\ref{eq.sigma}, \ref{eq.kappa}).
Indeed, to calculate $\sigma$ and $\kappa$ it was sufficient to approximate the tunneling matrix element $t_{pq}$ as a constant $t$ which is evaluated at the Fermi surface and neglect variations of $t_{pq}$ with energy which occur on the scale $T/\varepsilon_F$.
However, this approximation is insufficient for calculating the thermoelectric coefficient $\eta$ and thermopower $S$ since the dominant contribution to these quantities vanishes due to particle-hole symmetry such that both quantities are proportional to the small parameter $T/\varepsilon_F$.
Since it is necessary to take into account terms of order of $T/\varepsilon_F$ in order to obtain a nonzero result for $\eta$ and $S$, the corresponding expansions must be carried out to this order for all quantities which depend on energy.

For granular metals the important element of the diagram is the tunneling matrix
elements $t_{ij}^{kq}$ describing the coupling between grains $i$ and $j$. Therefore we derive an expression for $t_{ij}^{kq}$ in the following,
assuming that i and j are nearest neighbor grains and $t_{ij}^{kq}$ is independent
of the position in the sample. In order to calculate the energy dependence of these elements we
assume the tunneling barrier between grains to be a delta potential.
For the one-particle Hamiltonian $\widehat{H}=-\frac{\hbar ^{2}}{2m}\frac{d^{2}}{
dx^{2}}+\lambda \delta \left( x\right) $, with $\lambda$ being the strength of the potential,
the transmission rate for a single particle with energy $\varepsilon _{p}=\varepsilon _{F}+\xi _{p}$ is
\begin{eqnarray}
\label{tp}
T_{p} = \left( 1+\frac{m\lambda ^{2}}{2\hbar ^{2}\varepsilon _{p}}\right)
^{-1}\simeq \left( 1+\frac{m\lambda ^{2}}{2\hbar ^{2}\varepsilon _{F}}\left(
1-\xi _{p}/\varepsilon _{F}\right) \right) ^{-1} \\
 = T_{0}\left( 1+T_0^{-1}-\xi _{p}/\varepsilon _{F}\right) ^{-1} \nonumber\,.
\end{eqnarray}
Here $T_0 =\frac{m\lambda ^{2}}{2\hbar ^{2}\varepsilon _F}$ is the transmission rate at the Fermi level
and we use the fact that $\xi_p \ll \varepsilon _{F}$.
Next, we consider the case of large barriers, in this regime $T_{p}\simeq T_{0}\left( 1+\xi
_{p}/\varepsilon _{F}\right) $. In granular systems we have many channels
and have to consider tunneling processes with energy $\xi _{i}$ in grain $i$
and with $\xi _{j}$ in neighboring grain $j$: $t^{2}\propto N\left(
T_{p_{i}}^{2}+T_{p_{j}}^{2}\right) $. Thus, the final expression for tunneling matrix element is
\begin{equation}
\label{t}
t^{2}(\xi _{i},\xi _{j}) = t_{0}^{2}\left( 1+\frac{\xi _{i}+\xi _{j}}{%
\varepsilon _{F}}\right)\, ,
\end{equation}
where $t_0$ is the constant tunneling matrix element evaluated at the Fermi surface. For convenience we use $i=1$ and $j=2$ in the following.
Using Eq.~(\ref{t}) we obtain the following expression for the correlation function in Eq.~(\ref{eq.eta0grain})
\begin{subequations}
\begin{eqnarray}
\label{eq.Q0_def1}
Q^{(0)} &=&-set_{0}^{2}T \langle \overrightarrow{n}_{e}^{(0)}\cdot
\overrightarrow{n}_{h}^{(0)}\rangle \underset{\omega _{n}}{\sum }
a^{2d+2} \\
&&\times\int \frac{d^{d}p_{1}}{(2\pi )^{d}} \frac{d^{d}p_{2}}{(2\pi
)^{d}}\left[ \frac{\xi _{1}+\xi _{2}}{2}\right] \left[1+\frac{\xi _{1}+\xi
_{2}}{\varepsilon _{F}}\right] \nn\\
&&\times G(p_{1},\omega _{n})G(p_{2},\omega
_{n}+\Omega _{m})\,  \nonumber,
\end{eqnarray}
where $\overrightarrow{n}^{(0)}_{\alpha}$ is the unit vector in direction of the current $\alpha\in\{e,h\}$, $\langle \overrightarrow{n}_{e}^{(0)}\cdot
\overrightarrow{n}_{h}^{(0)}\rangle = 1/d$ is the result of averaging over angles, the summation goes over Fermionic Matsubara
frequencies $\omega_n = 2\pi T (n + 1/2)$, and $G(p,\omega_n)$ is the Matsubara Green's function
\begin{equation}
G(p,\omega _{n})=\left[ \imath \omega _{n}-\xi _{p}\pm \imath /(2\tau
_{\omega })\right] ^{-1},
\end{equation}
\end{subequations}
with $\xi _{p}=\varepsilon _{p}-\varepsilon _{F}$ being the electron energy
with respect to the Fermi energy [$\varepsilon_{p}=p^{2}/(2m)$] and the (energy)  relaxation time $\tau_{\omega}$ is defined in Eq.~(\ref{tau}). To shorten the notation, we neglect the momentum argument in the following and attach the grain index to the Green's function $G$.
The two $\xi$-factors under the integration in (\ref{eq.Q0_def1}) arise from the heat current, Eq.~(\ref{thermalvertex}), and the energy correction of the tunneling element, Eq.~(\ref{t}).
The momentum integrals in Eq.~(\ref{eq.Q0_def1}) are transformed into energy integrals taking into account first order corrections in $\xi/\varepsilon_F$ of the Jacobian [see Eq.~(\ref{transfomration}) of Appendix~\ref{app.hom}].

We first perform the analytical continuation over the Fermionic Matsubara frequencies $\omega
_{n}=2\pi T(n+1/2)$ in Eq.~(\ref{eq.Q0_def1}).
In order to accomplish that, the analytical
structure of diagram (a) in Fig.~\ref{fig.dia} needs to be analyzed, which gives rise to three different regions for the Matsubara summations:  $I_{1}=\left] -\infty ;-\Omega _{m}
\right] ,I_{2}=\left] -\Omega _{m};0\right[ ,I_{3}=\left[ 0;\infty \right[ $, in which
we can determine whether the Green's function is retarded, $G^-(\omega)$, or advanced, $G^+(\omega)$:
\begin{subequations}
\begin{eqnarray}
S_{1} &=& \underset{n\in I_{1}}{\sum }G_{1}^{-}(\omega
_{n})G_{2}^{-}(\omega _{n}+\Omega _{m}) \\
&=&\int \frac{-d\omega}{4\pi \imath T} \tanh \left( \frac{\omega}{2T}
\right) G_{1}^{-}(-\imath \omega +\imath \Omega )G_{2}^{-}(-\imath
\omega ),\nn \\
S_{2} &=&\underset{n\in I_{2}}{\sum }G_{1}^{+}(\omega _{n})G_{2}^{-}(\omega
_{n}+\Omega _{m}) \\
&=&\int \frac{-d\omega}{4\pi \imath T} \tanh \left( \frac{\omega}{2T}
\right) \left[ G_{1}^{-}(-\imath \omega )G_{2}^{+}(-\imath \omega
-\imath \Omega )\right.\nn\\
&& \left.-G_{1}^{-}(-\imath \omega +\imath \Omega )G_{2}^{+}(-\imath
\omega )\right],\nn \\
S_{3} &=& \underset{n\in I_{3}}{\sum }%
G_{1}^{+}(\omega _{n})G_{2}^{+}(\omega _{n}+\Omega _{m})\\
&=&\int \frac{-d\omega}{4\pi \imath T} \tanh \left( \frac{\omega}{2T}
\right) G_{1}^{+}(-\imath \omega )G_{2}^{+}(-\imath \omega -\imath
\Omega ).\nn
\end{eqnarray}
\end{subequations}
To calculate $\eta^{(0)}$  in Eq.~(\ref{eq.eta0grain}) we now consider the derivative of $S_1 + S_2 + S_3$ with respect to Bosonic frequencies $\left. \frac{\partial }{\partial \Omega }\right\vert _{\Omega =0}$.
For brevity we omit arguments $-\imath \omega $ of Green's functions, leading to
\begin{eqnarray}
 \left. \frac{\partial }{\partial \Omega }\right\vert _{\Omega =0}\left(
S_{1} + S_{2} + S_{3} \right) = \\
\int \frac{-d\omega}{4\pi \imath T} \tanh \left( \frac{\omega}{2T} \right) \nn
\frac{\partial }{\partial \omega }\left[ \frac{1}{\tau _{\omega }^{2}}
G_{1}^{-}G_{1}^{+}G_{2}^{-}G_{2}^{+}\right].\label{eq.divsum}
\end{eqnarray}
Next, one can perform the integration over variables $\xi_1$ and $\xi_2$
using Eqs.~(\ref{tau}) and (\ref{eq.divsum}) and the residuum theorem
\begin{subequations}
\begin{eqnarray}
&& \int d\xi _{1}d\xi _{2}\ g_{12}(\xi_1,\xi_2)G_{1}^{-}G_{1}^{+}G_{2}^{-}G_{2}^{+} \\
&&=  4\pi ^{2}\tau ^{2}\left[ 2\omega +\imath /\tau_0 +\frac{d}{2\varepsilon _{F}%
}\left( 2\omega +\imath /\tau_0 \right) ^{2}\right]\,, \nonumber
\end{eqnarray}
with
\begin{equation}
\label{g122}
g_{12}(\xi_1, \xi_2) = \xi _{1}+\xi _{2}+\frac{d}{2}\frac{\left( \xi
_{1}+\xi _{2}\right) ^{2}}{\varepsilon _{F}}.
\end{equation}
\end{subequations}
As a result we obtain the following expression for the derivative of correlation function
\begin{eqnarray}\label{eq.Q00}
\left. \frac{\partial }{\partial \Omega }\right\vert _{\Omega =0}Q^{(0)}
&=&-\frac{\pi s}{16\imath}et_{0}^{2}a^{2d+2}\left( \nu _{d}^{(0)}\right)
^{2} \frac{1}{T \varepsilon _{F}} \\
&&\times \int d\omega \frac{%
\left[ 2+(d/2-1)\imath /\left( \tau _{0}\varepsilon _{F}\right) \right]
^{2}\omega ^{2}}{\cosh ^{2}\left( \omega /(2T)\right) } \nonumber \\
&= &-\frac{\pi^3 s}{3 \imath }et_{0}^{2}a^{2d+2}\left( \nu
_{d}^{(0)}\right) ^{2} \frac{T^2}{\varepsilon _{F}}. \nonumber
\end{eqnarray}
Here all contributions of order $
1/\varepsilon^2_{F}$ or smaller are neglected in the final expression.
Substituting this result, Eq.~(\ref{eq.Q00}), into Eq.~(\ref{eq.eta0grain})
we finally obtain the following expression for the non-interacting thermoelectric
coefficient of granular metals
\begin{equation}
\label{eta01}
\eta ^{(0)}=-\frac{s\pi ^{3}}{3}et_{0}^{2}a^{d+2}\left( \nu
_{d}^{(0)}\right) ^{2}\frac{T}{\varepsilon _{F}}.
\end{equation}
One can re-write this expression using the relations: $\nu _{d}^{(0)}\mathcal{D}%
_{d}=g_T a^{2-d}$ ; \ $\nu _{d}^{(0)}=(\delta a^{d})^{-1}$ ; $t_{0}^{2}=g_T\delta
^{2}/(2\pi )$, where $\mathcal{D}_{d}$ is the diffusion constant, $g_T$ the
tunneling conductance, and $\delta $ the mean level spacing, giving
\begin{equation}\label{eq.eta0alt}
\eta^{(0)}= - \frac{s\pi^2}{6} eg_T a^{2-d} (T/\varepsilon _{F}).
\end{equation}

\subsection{correction due to Coulomb interaction}

Now, we consider the correction $\eta^{(1)}$ to the thermoelectric coefficient of granular metals due to electron-electron interaction
\begin{equation}
\eta=\eta^{(0)}+\eta^{(1)},
\end{equation}
where $\eta^{(0)}$ is given by Eq.~(\ref{eta01}).
Analogously to $\eta^{(0)}$, $\eta^{(1)}$ can be obtained from
\begin{equation}
\label{eta1sum1}
\eta ^{(1)}=\imath \left. \frac{\partial }{a^{d}T\partial \Omega }
\right\vert _{\Omega =0}\left( Q^{(1)}+Q^{(2)}+Q^{(3)}\right),
\end{equation}
where the diagrams $Q^{(1)}$, $Q^{(2)}$, $Q^{(3)}$ contributing to $\eta^{(1)}$ are shown in Fig.~\ref{fig.dia} (b, c, d).
Detailed calculations of $\eta^{(1)}$ are presented in Appendix~\ref{app.eta1}. However, here we outline the main steps of this derivation.
These three diagrams include the effect of elastic scattering of electron at impurities described by diffusons
\begin{subequations}
\begin{equation}
\mathcal{D}^{-1} = \tau _{\omega }\left( \left\vert \Omega _{i}\right\vert + \epsilon _{q}\delta \right),
\end{equation}
with
\begin{equation}
\epsilon _{q}=2g_{T}\left[ 2d-\sum_{a}^{\prime }\cos (
\overrightarrow{q}\cdot \overrightarrow{a})\right],
\end{equation}
\end{subequations}
where $\sum_{a}^{\prime }$ stands for summation over all directions and
orientations $\left\{ \pm a\overrightarrow{e}_{j}^{(0)}\right\}$,
and the effect of the dynamically screened Coulomb potential $\widetilde{V}(q,\Omega _{i}) = \mathcal{D}V(q,\Omega _{i}) \mathcal{D}$
\begin{subequations}
\begin{eqnarray}
\widetilde{V}(q,\Omega _{i}) &=& \frac{2E_{c}(q)}{\tau _{\omega }^{2}\left[ \left\vert \Omega
_{i}\right\vert +4E_{c}(q)\epsilon _{q}\right] \left[ \left\vert \Omega
_{i}\right\vert +\epsilon _{q}\delta \right] },  \\
V(q,\Omega _{i}) &=&\left( \frac{1}{2E_{c}(q)}+\frac{2\epsilon _{q}}{
\left\vert \Omega _{i}\right\vert +\epsilon _{q}\delta }\right) ^{-1}, \nn
\end{eqnarray}
where we use the notation
\begin{equation}\label{eq.Ecq}
E_{c}(q) = \frac{e^{2}}{a^{d}}\left\{
\begin{array}{l}
-\ln (qa), \hspace{0.4cm} d=1 \\
\pi /q,  \hspace{1cm} d=2 \\
2\pi /q^{2}, \hspace{0.7cm} d=3.
\end{array}
\right.
\end{equation}
\end{subequations}
Each diagram in Fig.~\ref{fig.dia} has also two types of renormalized interaction vertices: (i) the inter-grain vertex
\begin{eqnarray}
\label{Phi1}
\Phi _{\omega }^{(1)}(\Omega _{i}) =
\int
\frac{a^d d\overrightarrow{q}}{(2\pi )^{d}}
\frac{2E_{c}(q)\sum_{a}^{\prime }\cos (\overrightarrow{q}
\cdot \overrightarrow{a})}{\tau _{\omega }^{2}\left[ \left\vert \Omega_{i}
\right\vert +4E_{c}(q)\epsilon _{q}\right] \left[ \left\vert \Omega_{i}
\right\vert +\epsilon _{q}\delta \right] },
\end{eqnarray}
and (ii) the intra-grain vertex
\begin{eqnarray}
\label{Phi2}
\Phi _{\omega }^{(2)}(\Omega _{i}) = \int
\frac{ a^d d\overrightarrow{q}}{(2\pi )^{d}}\frac{2E_{c}(q)\, 2d}{\tau _{\omega }^{2}\left[
\left\vert \Omega _{i}\right\vert +4E_{c}(q)\epsilon _{q}\right] \left[
\left\vert \Omega _{i}\right\vert +\epsilon _{q}\delta \right] }\,.
\end{eqnarray}
Explicitly, the contribution $Q^{(1)}$ in Eq.~(\ref{eta1sum1}) [diagram (b) in Fig.~\ref{fig.dia}] is given by
\begin{subequations}
\begin{eqnarray}
\label{Q11def}
Q^{(1)} &=&-\frac{s}{2d}et_{0}^{2}T^{2}a^{2d+2}\left( \nu _{d}^{(0)}\right)
^{2} \\
&&\times \underset{\omega _{n},\Omega _{i}}\sum \int d\xi _{1}d\xi
_{2}g_{12} F_{1}^{\left( s_{1}s_{2}s_{3}s_{4}\right)
}\Phi _{\omega }^{(1)}\left( \Omega _{i}\right), \nn
\end{eqnarray}
where the function $g_{12}$ is defined in Eq.~(\ref{g122}) and we use the notation
\begin{eqnarray}
F_{1}^{\left( {s_{1}{s_{2}}}{s_{3}{s_{4}}}
\right) } &=& G_{1}^{s_{1}}(\omega _{n} + \Omega _{m})G_{1}^{s_{2}}(\omega
_{n}+\Omega _{m}+\Omega _{i}) \nonumber \\
&&\times G_{2}^{s_{3}}(\omega _{n})G_{2}^{s4}(\omega
_{n}+\Omega _{i}),
\end{eqnarray}
\end{subequations}
with $s_{i}=\pm$ denote the analytic structure of the Green's functions
implying restrictions on the frequency summation.

For the contribution $Q^{(2)}$ in Eq.~(\ref{eta1sum1}) [diagram (c) in Fig.~\ref{fig.dia}]
we have the following expression
\begin{subequations}
\begin{eqnarray}
\label{Q22}
Q^{(2)} &=& -\frac{s}{2d}et_{0}^{2}T^{2}a^{2d+2}\left( \nu _{d}^{(0)}\right)
^{2} \\
&&\times \underset{\omega _{n},\Omega _{i}}{\sum }\int d\xi _{1}d\xi
_{2}g_{12} F_{2}^{\left( {{s_{1}}{s_{2}}}s_{3}\right) }\Phi _{\omega}^{(2)}\left( \Omega _{i}\right),
\nonumber
\end{eqnarray}
where we use the notation
\begin{equation}
F_{2}^{\left( {{s_{1}}{s_{2}}}s_{3}\right) }=\left[
G_{1}^{s_{1}}(\omega _{n}+\Omega _{m})\right] ^{2}G_{1}^{s_{2}}(\omega
_{n}+\Omega _{m}+\Omega _{i})G_{2}^{s_{3}}(\omega _{n}).
\end{equation}
\end{subequations}

The diagram $Q^{(3)}$, shown in Fig.~\ref{fig.dia} (d), describes the contribution of the correlation function with the interaction part of the heat current operator, $\widehat{\jmath }_{ij}^{(h,1)}$ [second term in the right hand side of Eq.~(\ref{thermalvertex})], and has therefore a different structure in comparison with contributions $Q^{(1)}$ and $Q^{(2)}$:
\begin{subequations}
\begin{eqnarray}
Q^{(3)} &=&-\frac{s}{2d}et_{0}^{2}T^{2}a^{2d+2}\left( \nu _{d}^{(0)}\right)^{2} \\
&&\times \underset{\omega _{n},\Omega _{i}}{\sum }\int d\xi _{1}d\xi
_{2}g_{3} F_{3}^{\left( {s_{1}{s_{2}}}s_{3}\right) }\Phi _{3}\left(\Omega _{i},q\right), \nn
\end{eqnarray}
where
\begin{eqnarray}
F_{3}^{\left( {s_{1}{s_{2}}}s_{3}\right) }&=& G_{1}^{s_{1}}(\omega _{n}+\Omega _{m}+\Omega _{i})G_{1}^{s_{2}}(\omega_{n}+\Omega _{m})G_{2}^{s_{3}}(\omega _{n}), \nn \\
g_{3}&=& 2\left( 1+\frac{d}{2\varepsilon _{F}}\left( \xi _{1}+\xi
_{2}\right) \right).
\end{eqnarray}
\end{subequations}
The main contribution to $\eta^{(1)}$ from diagram $Q^{(3)}$ is of the order of $(T/\varepsilon_{F})^{2}$,
whereas  $Q^{(1)}$ and  $Q^{(2)}$ have $1/\varepsilon _{F}$ contributions.
Therefore we will not consider diagram $Q^{(3)}$ any further, but keep contributions of order $T/\varepsilon_{F}$ only.

\begin{figure}[t]
\includegraphics[width=0.8\columnwidth]{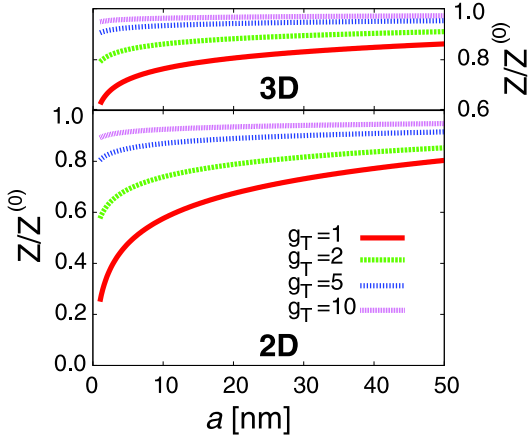}
\caption{(color online) Plots of the dimensionless figure of merit $Z/Z^{(0)}$  vs. grain size $a$ (in nm) - for different values of dimensionless tunneling conductance $g_T$ (see legend): the upper panel is for the three-dimensional (3D) case and the lower for the two-dimensional (2D). All curves are plotted for $T=100$K. At this temperature, the dimensionless bare figure of merit for granular metals is $Z^{(0)}T\approx 10^{-4}$.}
\label{fig.plot_Za}
\end{figure}

Thus, the first order interaction corrections to the thermoelectric coefficient are only generated by diagrams (b) and (c) in Fig.~\ref{fig.dia}. Substituting Eqs.~(\ref{Q11def}) and (\ref{Q22}) into Eq.~(\ref{eta1sum1}) after summation over Fermionic, $\omega_n$, and Bosonic, $\Omega_i$, frequencies and analytical continuation we obtained (see Appendix~\ref{app.eta1} for details)
\begin{equation}
\label{eta1}
\eta ^{(1)} = -\frac{\eta ^{(0)}}{2\pi g_{T}}\left( \frac{a}{2\pi }\right) ^{d}\int d^d {\bf q} \ln \left[\frac{2E_c({\bf q})\epsilon_{{\bf q}}}{T}\right],
\end{equation}
where the ${\bf q}$-integration goes over the $d$-dimensional sphere with radius $\pi/a$.
Integrating over ${\bf q}$ in Eq.~(\ref{eta1}) we obtain the following expressions, neglecting all constants under the logarithm, in two ($2D$) and three ($3D$) dimensions:
\begin{equation}
\eta _{2D}^{(1)} = -\frac{\eta ^{(0)}}{8 g_{T}}
\ln \frac{E_c g_{T}}{T}, \hspace{0.5cm}
\eta _{3D}^{(1)} = -\frac{\eta ^{(0)}}{12 g_{T}}\ln \frac{E_c g_{T}}{T},
\end{equation}
which lead to Eq.~(\ref{eq.eta}).

\section{Thermopower and Figure of Merit}\label{sec.SZ}

\begin{figure}[t]
\includegraphics[width=0.9\columnwidth]{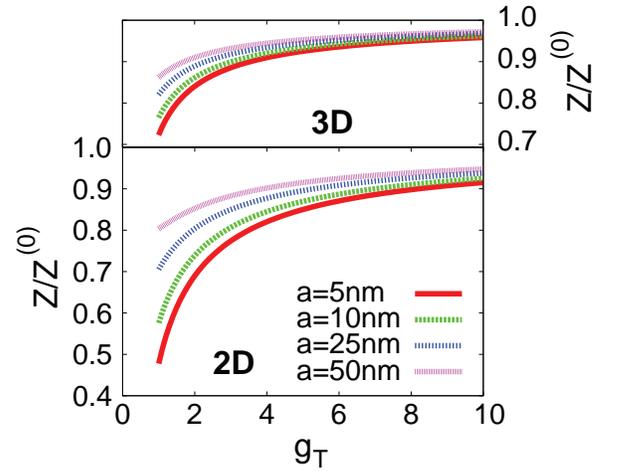}
\caption{(color online) Plots of the dimensionless figure of merit $Z/Z^{(0)}$  vs. tunneling conductance $g_T$ - for different values of grain sizes $a$ (see legend): the upper panel is for the three-dimensional (3D) case and the lower for the two-dimensional (2D). All curves are plotted for $T=100$K.}
\label{fig.plot_Zg}
\end{figure}

Now we have the expressions for all three kinetic coefficients $\sigma$, $\kappa$, and $\eta$ for granular metals to order $T/\varepsilon_F$.
Based on these, we can derive other thermodynamic quantities -- in particular we discuss the thermopower (Seebeck coefficient) and figure of merit in this section.  Both quantities are relevant parameters for thermocouples, the thermopower is a measure the change of voltage due to a temperature gradient and the figure of merit a measure for the performance of the device.

The thermopower is related to the kinetic coefficients as (see Eq.~(\ref{00}))
\begin{subequations}
\begin{equation}\label{eq.S_def}
S=\eta/\sigma\,.
\end{equation}
Again, we only consider terms up to order $T/\varepsilon_F$ and obtain the expression
\begin{equation}
S = \frac{\eta^{(0)}}{\sigma^{(0)}}\left(\frac{\eta^{(1)}}{\eta^{(0)}}-\frac{\sigma^{(1)}}{\sigma^{(0)}}\right)\,,\label{eq.S_result}
\end{equation}
\end{subequations}
which results in Eq.~(\ref{eq.S}).

The dimensionless figure of merit is related to the kinetic coefficients as
\begin{subequations}
\begin{equation}\label{eq.Z_def}
ZT =T\eta^2/(\sigma\kappa)= S^2\sigma T/\kappa,
\end{equation}
giving
\begin{equation}\label{eq.Z_result}
ZT = \frac{(S^{(0)})^2\sigma^{(0)} T}{\kappa^{(0)}}\left(  1- \frac{\kappa^{(1)}}{\kappa^{(0)}} + \frac{ \sigma^{(1)}}{\sigma^{(0)}} + 2\frac{S^{(1)}}{S^{(0)}} \right),
\end{equation}
\end{subequations}
resulting in Eq.~(\ref{eq.ZT}), which has lowest order $(T/\varepsilon_F)^2$ [Eq.~\ref{eq.Z0}].
In Eq.~(\ref{eq.Z_result}) $\kappa^{(1)}$, $\sigma^{(1)}$, and $S^{(1)}$ are corrections to the thermal conductivity, electrical conductivity, and the Seebeck coefficient due to Coulomb interaction, respectively.
The numerical coefficient two in front of the last term reflects the fact that the Seebeck coefficient appears squared in the definition of $ZT$.
Using Eqs.~(\ref{eq.S}) and (\ref{eq.Z_result}) one can see that the second term of the right-hand-side of Eq.~(\ref{eq.ZT}) originates due to correction to the Seebeck coefficient, Eq.~(\ref{eq.S}).
The origin of the third term in Eq.~(\ref{eq.ZT}) is due to correction to the thermal conductivity, $\kappa$, last term on the right hand side of  Eq.~(\ref{eq.kappa}).
We mention that the second term on the right hand side of Eq.~(\ref{eq.kappa}) cancels with the correction to the electrical conductivity, Eq.~(\ref{eq.sigma}), after the substitution into Eq.~(\ref{eq.Z_result}).

\begin{figure}[t]
\includegraphics[width=0.8\columnwidth]{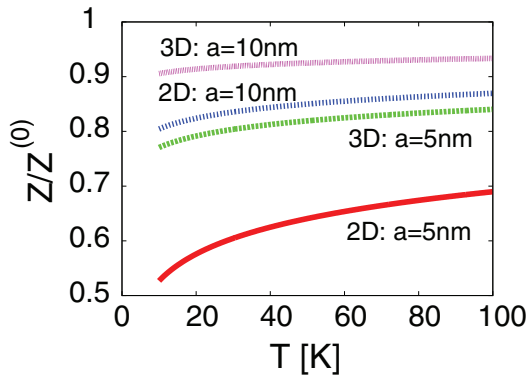}
\caption{(color online) Plots of the dimensionless figure of merit $Z/Z^{(0)}$  vs. temperature $T$ for $g_T=5$ (upper two graphs) and $g_T=2$ (lower two graphs) and grain size $a=5,10$nm and $d=2,3$. Legends are next to the corresponding curves.}
\label{fig.plot_ZT}
\end{figure}
In Fig.~\ref{fig.plot_Za} the dependence of $Z$ on the grain size $a$ and in Fig.~\ref{fig.plot_Zg} on the tunneling conductance $g_T$ for two- and three-dimensional samples are shown. Fig.~\ref{fig.plot_ZT} shows the temperature dependence of the figure of merit.
These plots show that the correction term to $ZT$ is most effective for small grain at not very high tunneling conductance and at low temperatures.

\section{Discussions}
\label{sec.discussions}

In the presence of interaction effects and not very low temperatures $T > g_T \delta$, granular metals behave differently from homogeneous disorder metals. However, in the absence of interactions the result for $\eta^{(0)}$ below Eq.~(\ref{eq.eta}) [or Eq.~\ref{eq.eta0alt}] coincides with the thermoelectric coefficient of homogeneous disordered metals (see Appendix~\ref{app.hom}),
\begin{equation}
\eta^{(0)}_{hom} = -(2/9)e p_F (\tau_0 T),
\end{equation}
with $p_F$ being the Fermi momentum. One can expect that at low  temperatures, $T < g_T \delta$, even in the presence of Coulomb interaction the behavior of thermoelectric coefficient and thermopower of granular metals is similar to the behavior of $\eta_{hom}$ and $S_{hom}$, however this temperature range is beyond the scope of the present paper.
Our results for thermopower (\ref{eq.S}) and figure of merit (\ref{eq.ZT}) show that the influence of Coulomb interaction is most effective for small grains. The thermopower $S^2$ decreases with the grain size which is a result of the delicate competition of the corrections of thermoelectric coefficient (\ref{eq.eta}) and the electric conductivity (\ref{eq.sigma}). In particular, if the numerical prefactor of the correction to $\eta$ would be slightly smaller, the sign of the correction to $S$ would change.

Above we only considered the electron contribution to the figure of merit. At higher temperatures $T>T^*$, where
\begin{equation}
T^* \sim \sqrt{g_T c_{ph}^2/l_{ph}} a,
\end{equation}
is a characteristic temperature with  $l_{ph}$ and $c_{ph}$ being the phonon scattering length and phonon velocity respectively~\cite{Beloborodov05}, phonons will provide an independent, additional contribution to thermal transport,
\begin{equation}
\kappa_{ph} = T^3 l_{ph}/c_{ph}^2.
\end{equation}
However, the phonon contribution can be neglected for temperatures
\begin{equation}
g_T \delta < T < T^*.
\end{equation}
A detailed study of the influence of phonons at high temperatures, including room temperature, will be subject of a forthcoming work.

So far, we ignored the fact that electron-electron interactions also renormalize the chemical
potential $\mu$. In general, this renormalization may affect the kinetic coefficients: the thermal current vertex, Fig~\ref{fig.jver}, as well as the electron Green's functions depend on $\mu$. In particular one needs to replace $\nabla (e V) \rightarrow \nabla (e V + \mu)$ in Eq.~(\ref{responce}). To first order in the interactions, the renormalization of $\mu$ only leads to corrections to diagram (a) in Fig.~\ref{fig.dia}. As it can be easily shown, for this diagram the renormalization of the heat and electric current vertices is exactly canceled by the renormalization of the two electron propagators. Therefore, the renormalization of the chemical potential does not affect our results in the leading order.

Finally, we remark that the bare figure of merit $Z^{(0)}T$ for granular metals at $g_T > 1$ and $100$K is of the order of only $10^{-4}$. Therefore these materials are not suitable for {\it solid-state refrigerators}, but should be replaced by granular semiconductors with $g_T < 1$.
However, the case of granular metals is still relevant for low temperature applications in, e.g., thermocouples.
Therefore we conclude this paragraph discussing the dimensionless figure of merit $ZT$ of granular materials at weak coupling between the grains, $g_T \ll 1$. In this regime the electronic contribution to thermal conductivity $\kappa_e$ of granular metals was recently investigated in Ref.~\onlinecite{Tripathi}, where it was
shown that
\begin{equation}
\kappa_e \sim g_T^2 T^3/E_c^2.
\end{equation}
In this regime the electric conductivity of granular metals obeys the law~\cite{Efetov02,Beloborodov07}
\begin{equation}
\sigma \sim g_T \exp(-E_c/T).
\end{equation}
However, an expression for the thermoelectric coefficient in this region is not available yet, but recently it has been proposed, based on experiment, that nanostructured thermoelectric materials in the low coupling region (AgPb$_m$SbT$e_{2+m}$, Bi$_2$Te$_3$/Sb$_2$Te$_3$, or CoSb$_3$)~\cite{Hsu,Poudel,Majumdar,Mi,harman02} can have higher figures of merit than their bulk counterparts.

In conclusion, we have investigated the thermoelectric coefficient and thermopower of granular nanomaterials in the limit of large tunneling conductance between the grains and temperatures $T > g_T\delta$. We have shown to what extend quantum and confinement effects in granular metals are important in changing $ZT$ depending on system parameters. We also presented the details of our calculations.

\paragraph*{Acknowledgements} We thank Frank Hekking, Nick Kioussis, and Gang Lu for
useful discussions. A.~G. was supported by the
U.S. Department of Energy Office of Science under the Contract No. DE-AC02-06CH11357.

\appendix

\section{Thermoelectric coefficient of homogeneous disordered metals in the absence of interaction}
\label{app.hom}

\begin{figure}[t]
\includegraphics[width=0.8\linewidth]{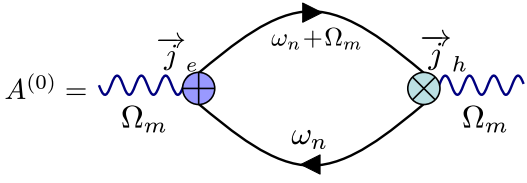}
\caption{(color online) Lowest order diagram for the heat-electric current correlator for homogeneous disordered metals. The external Bosonic frequency is denoted by $\Omega$ (wavy lines) and the internal Fermionic frequency by $\omega$ (straight lines).
The electric and heat current vertexes are $\protect\overrightarrow{\jmath_e}$ and $\protect\overrightarrow{\jmath_h}$, respectively.}\label{fig.A0}
\end{figure}
In order to demonstrate important steps of our calculations, we present a derivation of the thermoelectric coefficient for homogeneous disordered metals in the absence of interaction in this appendix.
In linear response theory the thermoelectric coefficient can be written as
\begin{equation}
\eta ^{(0)}=\imath \left. \frac{\partial }{L^{d}T\partial \Omega }%
\right\vert _{\Omega =0}A^{(0)}\,,
\end{equation}
where the diagrammatic representation of the correlator $A^{(0)}$ is shown in Fig.~\ref{fig.A0}
with electric and heat current vertexes
\begin{equation}
\overrightarrow{\jmath _{e}}=\frac{e}{m}\underset{p}{\sum }
\overrightarrow{p}{\widehat{a}}_{p}^{\dag}{\widehat{a}}_{p},
\hspace{0.5cm} \overrightarrow{\jmath _{h}}=\frac{1}{m}
\underset{p}{\sum }\overrightarrow{p}\xi _{p}{\widehat{a}}_{p}^{\dag}{\widehat{a}}_{p},
\end{equation}
respectively. The sums over momentum are transformed into integrals $\underset{p}{\sum }\rightarrow \left( \frac{L}{2\pi }\right) ^{d}\int d^{d}p$
by which we can transform the sum over the momentum product to $\int d^{d}p_{e}\int
d^{d}p_{h}\overrightarrow{p}_{e}\cdot \overrightarrow{p}_{h}=\int
d^{d}p_{e}\int d^{d}p_{h}\left\vert \overrightarrow{p}_{e}\right\vert
^{2}\left\langle \overrightarrow{n}_{e}\cdot \overrightarrow{n}%
_{h}\right\rangle \delta \left( \overrightarrow{p}_{e}-\overrightarrow{p}%
_{h}\right) \left( \frac{2\pi }{L}\right) ^{d}$, where $\overrightarrow{n}%
_{\alpha}$ is the unit vector in direction of the current $\alpha\in\{e,h\}$.
Averaging over angles gives
$\left\langle \overrightarrow{n}_{e}\cdot \overrightarrow{n}_{h}\right\rangle
=1/d$, and therefore:
\begin{eqnarray}
\nonumber
A^{(0)}(k,\Omega _{m})= -\frac{s}{d}\frac{e}{m^{2}}\underset{\omega _{n}}{T\sum }L^{d}\int
\frac{d^{d}p}{(2\pi )^{d}}\left\vert \overrightarrow{p}\right\vert ^{2}  \\
\times \xi_{p}G(p+k,\omega _{n}+\Omega _{m})G(p,\omega _{n}),
\label{A0}
\end{eqnarray}
where $s$ is the spin degeneracy factor, $\Omega _{m}$ an external bosonic Matsubara frequency, and $G(p,\omega_n)$ the momentum dependent Matsubara Green's function.
In the following we only consider the case of zero external momentum, $k=0$.
For the {\it advanced} (in $\mathbb{C}^{+}$) and {\it retarded} (in $\mathbb{C}^{-}$) Green's function, we use
\begin{equation}
\label{GF}
G^{\pm }(p,\omega _{n})=\left[ \imath \omega _{n}-\xi _{p}\pm \imath /(2\tau
_{\omega })\right] ^{-1},
\end{equation}
where $\xi _{p}=\varepsilon _{p}-\varepsilon _{F}$ is the electron energy
with respect to the Fermi energy [$\varepsilon
_{p}=p^{2}/(2m)$] and the (energy)  relaxation time $\tau _{\omega }$ depends on
the (real) frequency $\omega $.

The momentum integral in Eq.~(\ref{A0}) is transformed into an energy integral as follows
\begin{equation}
\int \frac{d^{d}p}{(2\pi )^{d}}f(|\overrightarrow{p}|)
=\frac{\Omega_d 2^{\frac{d}{2}-1}m^{\frac{d}{2}}}{(2\pi )^{d}}\int_{0}^{\infty }f(
\sqrt{2m\varepsilon }) \varepsilon ^{\frac{d}{2}-1}d\varepsilon,
\label{pintegral}
\end{equation}
where $\Omega_d$ is the value of the angular integral ($\Omega_{1,2,3}=\{2,2\pi ,4\pi \}$).
Since the Green's function depends on $\xi$, we need to rewrite this
integral using the $d\varepsilon =d\xi $ and $\varepsilon ^{d/2-1}=\varepsilon
_{F}^{d/2-1}\left( 1+\xi /\varepsilon _{F}\right) ^{d/2-1}$
\begin{eqnarray}
\label{eintegral}
\int_{0}^{\infty }\varepsilon ^{d/2-1}d\varepsilon &=&\varepsilon
_{F}^{d/2-1}\int_{-\varepsilon _{F}}^{\infty }\left( 1+\xi /\varepsilon
_{F}\right) ^{d/2-1}d\xi \\ \nn
&\approx& \varepsilon _{F}^{d/2-1}\int_{-\infty
}^{\infty }\left[ 1+(d/2-1)\xi /\varepsilon _{F}\right] d\xi\,.
\end{eqnarray}
Combining Eqs.~(\ref{pintegral}) and (\ref{eintegral}) we obtain
\begin{eqnarray}
\label{transfomration}
\int \frac{d^{d}p}{(2\pi )^{d}}f(|\overrightarrow{p}|)&\approx& \nu
_{d}^{(0)}\int_{-\infty }^{\infty }\left[ 1 + ( \frac{d}{2} - 1)\frac{\xi}{\varepsilon _{F}}%
\right] \\ \nn
 &&\times f\left( \sqrt{2m(\xi -\varepsilon _{F})}\right) d\xi,
\end{eqnarray}
with $\nu _{d}^{(0)}=\frac{2^{d/2-1}\Omega (d)}{(2\pi )^{d}}
m^{d/2}\varepsilon _{F}^{d/2-1}$ being the density of states (DOS) at the Fermi surface.

The DOS (and $\tau _{\omega }$) depends on $\omega $ via
\begin{eqnarray}
\label{01}
\nu _{d}(\omega )&=&\frac{1}{\pi }\int \frac{d^{d}p}{(2\pi )^{d}}\Im
G^{-}\\ \nn
&=&\frac{\nu _{d}^{(0)}}{\pi }\left[ 1 + \frac{\omega}{\varepsilon _F}\right]^{\frac{d}{2}-1}\int
\frac{(1+x/\widetilde{x})^{d/2-1}}{x^{2}+1}dx\,,
\end{eqnarray}
with $\widetilde{x}=2\tau _{\omega }\varepsilon _{F}(1+\omega /\varepsilon
_{F})$ and the symbol $\Im$ stands for imaginary part.
However, the term $x/\widetilde{x}$ is neglected in the integral,
hence [$\int_{\mathbb{R}} \left( x^{2}+1\right) ^{-1}dx=\pi $]
\begin{equation}
\label{02}
\tau _{\omega }^{-1}=\tau _{0}^{-1}\frac{\nu _{d}(\omega )}{
\nu _{d}^{(0)}}\approx \tau _{0}^{-1}\left[ 1+(d/2-1)\omega /\varepsilon _{F}
\right].
\end{equation}
For conveniens we drop the momentum argument of $G$ in the following, as well as the $-\imath\omega$ argument.

Using all the above equations we finally obtain
\begin{eqnarray}
\eta ^{(0)} &=& - \frac{2s}{d}\frac{\imath e}{m}\nu _{d}^{(0)}\left. \frac{
\partial }{\partial \Omega }\right\vert _{\Omega =0}\underset{\omega _{n}}{
\sum }\int d\xi  \\ \nn
&&\times \left[ 1+ (d/2-1)\xi /\varepsilon _{F}\right] \xi (\xi
-\varepsilon _{F}) \\ \nn
&&\times \frac{1}{\imath (\omega _{n}+\Omega _{m})-\xi +\imath \sgn
(\omega _{n}+\Omega _{m})/(2\tau )}\\ \nn
&&\times\frac{1}{\imath \omega _{n}-\xi +\imath
\sgn(\omega _{n})/(2\tau )}\,.
\end{eqnarray}
Here the derivative by the real external frequency, $\Omega$, requires the analytic continuation of the Matsubara expression.
For the following calculation of the $\omega_n$ sum and energy integral, it is convenient to introduce the notation
\begin{eqnarray}
g_{\xi } &\equiv& \left[ 1+(d/2-1)\xi
/\varepsilon _{F}\right] \xi (\xi -\varepsilon _{F}) \\
&=& -\xi \varepsilon
_{F}+(2-d/2)\xi ^{2}+O\left[ \xi ^{3}/\varepsilon _{F}\right] \nonumber.
\end{eqnarray}
Next, we split the sum over the Fermionic Matsubara frequencies $\omega
_{n}=2\pi T(n+1/2)$ into three intervals $I_{1}=\left] -\infty ;-\Omega _{m}
\right] ,I_{2}=\left] -\Omega _{m};0\right[ ,I_{3}=\left[ 0;\infty \right[ $ such that we can specify the analytical structure of the Green's function, i.e.
\begin{eqnarray}
\eta ^{(0)} && = -\frac{2s}{d}\frac{\imath e}{m}\nu _{d}^{(0)}\left. \frac{
\partial }{\partial \Omega } \right\vert_{\Omega =0}\int d\xi \, g_{\xi } \nn\\
&& \times \left[ \underset{n\in I_{1}}{\sum }G^{-}(\omega _{n} + \Omega_{m})
G^{-}(\omega _{n})\right. \nn\\
&& + \underset{n\in I_{2}}{\sum }G^{+}(\omega
_{n} + \Omega _{m})G^{-}(\omega _{n}) \nn\\
&& + \left. \underset{n\in I_{3}}{\sum }G^{+}(\omega _{n}+\Omega
_{m})G^{+}(\omega _{n})\right]  \nn\,.
\end{eqnarray}
Now, we perform the analytical continuation of the sums to real frequencies and at the same time of the external frequency:
\begin{eqnarray}
S_{1}&=&\underset{n\in I_{1}}{\sum }G^{-}(\omega _{n}+\Omega _{m})G^{-}(\omega
_{n})\\ \nn
&=& \int\limits_{\mathbb{R}}\frac{-d\omega}{4\pi \imath T} \tanh (\frac{\omega }{2T})G^{-}(-\imath \omega )G^{-}(-\imath \omega
+\imath \Omega )\,,
\end{eqnarray}
\begin{eqnarray}
S_{2}&=&\underset{n\in I_{2}}{\sum }G^{+}(\omega _{n}+\Omega _{m})G^{-}(\omega
_{n})  = \int\limits_{\mathbb{R}
}\frac{-d\omega}{4\pi \imath T} \tanh (\frac{\omega}{2T}) \\
 &\times& \left[ G^{+}(-\imath \omega -\imath \Omega
)G^{-}(-\imath \omega )-G^{+}(-\imath \omega )G^{-}(-\imath \omega +\imath
\Omega )\right], \nonumber
\end{eqnarray}
\begin{eqnarray}
S_{3}&=&\underset{n\in I_{3}}{\sum }G^{+}(\omega _{n}+\Omega _{m})G^{+}(\omega
_{n})\\ \nn
&=& \int\limits_{\mathbb{R}
}\frac{d\omega}{4\pi \imath T} \tanh (\frac{\omega}{2T})G^{+}(-\imath \omega -\imath \Omega
)G^{+}(-\imath \omega ).
\end{eqnarray}
Next, we rewrite the $\Omega $-derivative of the integrant in terms of
$\omega $-derivatives, (arguments $-\imath \omega $ are omitted):
\begin{eqnarray}
\label{derivative1}
\left. \frac{\partial }{\partial \Omega }\right\vert _{\Omega =0}G(-\imath \omega \mp \imath \Omega ) &=& \pm \frac{\partial }{\partial \omega }
G\,,
\end{eqnarray}
which is valid for both, advanced and retarded functions, and therefore we can simplify
\begin{widetext}
\begin{eqnarray}
\label{derivative2}
\left. \frac{\partial }{\partial \Omega }\right\vert _{\Omega =0}\left[
-G^{+}(-\imath \omega -\imath \Omega )G^{-}+G^{+}G^{-}(-\imath \omega
+\imath \Omega )\right] &=&\left( G^{+}\right) ^{2}G^{-}+G^{+}\left(
G^{-}\right) ^{2}=-\frac{\partial }{\partial \omega }\left( G^{+}G^{-}\right)\,,
\\
\left. \frac{\partial }{\partial \Omega }\right\vert _{\Omega =0}\left[
-G^{-}G^{-}(-\imath \omega +\imath \Omega )\right] &=&-G^{-}\left(
G^{-}\right) ^{2}=\frac{1}{2}\frac{\partial }{\partial \omega }\left(
G^{-}\right) ^{2}\,,  \notag \\
\left. \frac{\partial }{\partial \Omega }\right\vert _{\Omega =0}\left[
G^{+}(-\imath \omega -\imath \Omega )G^{+}\right] &=&-\left( G^{+}\right)
^{2}G^{+}=\frac{1}{2}\frac{\partial }{\partial \omega }\left( G^{+}\right)
^{2}\,. \nonumber
\end{eqnarray}
\end{widetext}
Using Eqs.~(\ref{derivative1}) and (\ref{derivative2}) we can write the following
\begin{eqnarray}
&&\left. \frac{\partial }{\partial \Omega }\right\vert _{\Omega =0}\left[
S_{1}+S_{2}+S_{3}\right]  =\nn\\
&& \int\limits_{\mathbb{R}} \frac{d\omega}{8\pi \imath T} \tanh \frac{\omega}{2T}   \frac{\partial }{\partial \omega }\left\{
(G^{+}-G^{-})^{2}\right\}\,.
\end{eqnarray}
We notice that at this point the $\omega $-integration should not be done by parts,
since the boundary term is important.
We can now do the $\xi $-integration using the fact that $G^{+}-G^{-}=-(\imath /\tau )G^{+}G^{-}$
\begin{eqnarray}
\label{xiintegration}
\Xi (\omega )&\equiv&\int d\xi \ g_{\xi }\left(\frac{\imath G^{+}G^{-}}{\tau}\right)^{2}\\
 &=&-\int \frac{g_{\xi }\,d\xi}{\left(\tau\left[ \omega -\xi +\imath /(2\tau )\right] \left[
\omega -\xi -\imath /(2\tau )\right] \right)^2} \nn\\
&=&-2\pi \left[ \tau \omega ^{2}(4-d)+\frac{4-d}{4\tau }-2\omega \tau
\varepsilon _{F}\right]\,.\nn
\end{eqnarray}
Here, the higher order terms in $g_{\xi }$ \ are neglected. In Eq.~(\ref{xiintegration}) we need
to keep the terms proportional to $\omega^{2}$ only, therefore
after $\omega $-expansion of $\tau=\tau_\omega $ we obtain:
\begin{equation}
\Xi (\omega )=-2\pi \left[ \tau _{0}\omega ^{2}(4-d)+\omega ^{2}\tau
_{0}(d-2)\right] =-4\pi \tau _{0}\omega ^{2}.
\end{equation}
And, finally
\begin{eqnarray}
\eta ^{(0)} &=& -\frac{s}{d}\frac{\imath e \nu _{d}^{(0)}}{m} \int\limits_{\mathbb{R}
}\frac{d\omega}{4\pi \imath T} \tanh (\frac{\omega}{2T})\frac{\partial }{\partial \omega }\Xi (\omega )
\nonumber \\
&=& - \frac{s}{2d}\frac{e}{m}\nu _{d}^{(0)} \frac{\tau _{0}}{T^2} \int\limits_{
\mathbb{R}
}\frac{\omega ^{2}d\omega }{\cosh ^{2}(\omega /2T)} \nonumber  \\
&=& - \frac{2 \pi^2 s}{3d}\frac{e}{m} \nu _{d}^{(0)}(\tau
_{0}T).
\end{eqnarray}
To perform the last integration we used the integral $\int \frac{x^{2}dx}{\cosh ^{2}(x)}=\pi ^{2}/6$.
In $d=3$ the density of states has the form $\nu_{d=3}^{(0)}=\frac{mp_{F}}{2\pi ^{2}}$, leading to
\begin{equation}
\eta _{3D}^{(0)} = -\frac{s}{9}ep_{F}(\tau _{0}T)\,.
\end{equation}

\section{Heat and electric current operators of granular metals}
\label{app.cur}

In this Appendix we derive an expression for the heat and electric current operators of granular metals in the presence of Coulomb interaction. The Hamiltonian for the granular system is $\widehat{\mathcal{H}}=\underset{i}{\sum }\widehat{\epsilon }_{i}$ where
\begin{eqnarray}
\label{b1}
\widehat{\epsilon }_{i} &=&\underset{k}{\sum }\xi _{k}\widehat{a}
_{i,k}^{\dagger }\widehat{a}_{i,k}+\frac{e^{2}}{2}\underset{j}{\sum }
\widehat{n}_{i}C_{ij}^{-1}\widehat{n}_{j} \\
&& + \frac{1}{2}\underset{j,k,q}{\sum }
\left[ t_{ij}^{kq}\widehat{a}_{i,k}^{\dagger }\widehat{a}_{j,q}+t_{ji}^{qk}
\widehat{a}_{j,q}^{\dagger }\widehat{a}_{i,k}\right] \nonumber \\
&\equiv &\widehat{\epsilon }_{i}^{(e)}+\widehat{\epsilon }_{i}^{(c)}+
\widehat{\epsilon }_{i}^{(t)}. \nonumber
\end{eqnarray}
Here we introduce the notation $\widehat{n}_{i}\equiv \underset{k}{\sum }\widehat{a}
_{i,k}^{\dagger }\widehat{a}_{i,k}$ for the number of electron within a grain $i$. The creation
(annihilation) operators $\widehat{a}
_{i,k}^{\dagger } (\widehat{a}_{i,k})$  satisfy the anti-commutation relations $\left\{ \widehat{a}_{i,k}^{\dagger };
\widehat{a}_{j,q}\right\} _{+}=\delta _{ij}\delta _{kq}$ and $\left\{
\widehat{a}_{i,k}^{(\dagger )};\widehat{a}_{j,q}^{(\dagger )}\right\} _{+}=0$.

The electric current through grain $i$ is
\begin{equation}
\label{b2}
-\imath \frac{d\widehat{n}
_{i}}{dt}=\left[ \widehat{n}_{i};\widehat{\mathcal{H}}\right] \equiv -\frac{\imath }{e}
\underset{j}{\sum }\widehat{\jmath }_{ij}^{(e)}.
\end{equation}

And the heat current through grain $i$ is
\begin{equation}
\label{b3}
-\imath \frac{d\widehat{\epsilon }_{i}}{dt}=\left[
\widehat{\epsilon }_{i};\widehat{\mathcal{H}}\right] \equiv -\imath \underset%
{j}{\sum }\widehat{\jmath }_{ij}^{(h)}.
\end{equation}
Accordingly the total current operators are given by
\begin{equation}
\label{b4}
\overrightarrow{\jmath }
_{e,h}=\overrightarrow{n}_{e,h}^{(0)}\sum\limits_{i,j}\widehat{\jmath }
_{ij}^{(e,h)},
\end{equation}
where $\overrightarrow{n}_{e,h}^{(0)}$ are the unit directions of the current (electric field, $E$, gradient) or heat (temperature, $T$, gradient) flows.

\subsection{electric current operator}
First, we calculate the electric current. Since the number operator of electrons $\widehat{n}_{i}$ commutes
with the first two terms of $\widehat{
\mathcal{H}}$ in Eq.~(\ref{b2}), i.e. $\left[ \widehat{n}_{i};\widehat{\mathcal{H}}\right] =\left[
\widehat{n}_{i};\widehat{\mathcal{H}}^{(t)}\right]$, we only need to calculate
\begin{widetext}
\begin{eqnarray}
\left[ \widehat{n}_{i};\widehat{\mathcal{H}}^{(t)}\right] &=&\frac{1}{2}%
\underset{k,i^{\prime },j,k^{\prime },q}{\sum }\left[ \widehat{a}%
_{i,k}^{\dagger }\widehat{a}_{i,k}\left( t_{i^{\prime }j}^{k^{\prime }q}%
\widehat{a}_{i^{\prime },k^{\prime }}^{\dagger }\widehat{a}%
_{j,q}+t_{ji^{\prime }}^{qk^{\prime }}\widehat{a}_{j,q}^{\dagger }\widehat{a}%
_{i^{\prime },k^{\prime }}\right) -\left( t_{i^{\prime }j}^{k^{\prime }q}%
\widehat{a}_{i^{\prime },k^{\prime }}^{\dagger }\widehat{a}%
_{j,q}+t_{ji^{\prime }}^{qk^{\prime }}\widehat{a}_{j,q}^{\dagger }\widehat{a}%
_{i^{\prime },k^{\prime }}\right) \widehat{a}_{i,k}^{\dagger }\widehat{a}%
_{i,k}\right]\nonumber \\
&=&\underset{j,k,q}{\sum }\left( t_{ij}^{kq}\widehat{a}_{i,k}^{\dagger }
\widehat{a}_{j,q}-t_{ji}^{qk}\widehat{a}_{j,q}^{\dagger }\widehat{a}
_{i,k}\right) \equiv \widehat{\mathcal{C}}_{i}^{(nt)} \label{eq.Cnt}
\end{eqnarray}
\end{widetext}
As a result we get the following expression for the electric current operator
\begin{equation}\label{je}
\widehat{\jmath }_{ij}^{(e)}=\imath e\underset{k,q}{\sum }
\left( t_{ij}^{kq}\widehat{a}_{i,k}^{\dagger }\widehat{a}_{j,q}-t_{ji}^{qk}
\widehat{a}_{j,q}^{\dagger }\widehat{a}_{i,k}\right)\,.
\end{equation}

\subsection{heat current operator}

Second we turn to the heat current, for which we need to calculate nine commutators (see Eq.~(\ref{b3}))
\begin{equation}
\label{b7}
\widehat{\mathcal{C}}_{i}^{(\alpha \beta )}\equiv \left[ \widehat{\epsilon}_{i}^{(\alpha )};\widehat{\mathcal{H}}^{(\beta )}\right],
\end{equation}
where $\alpha \in \{e,c,t\}$ and operator $\widehat{\epsilon}_{i}^{(\alpha )}$ was defined in Eq.~(\ref{b1}).

Since operators $\widehat{n}_{i}$ commute with each other, four of the commutators in Eq.~(\ref{b7}) vanish:
\begin{equation}
\widehat{\mathcal{C}}_{i}^{(ee)}=\widehat{
\mathcal{C}}_{i}^{(cc)}=\widehat{\mathcal{C}}_{i}^{(ec)}=\widehat{\mathcal{C}
}_{i}^{(ce)}=0.
\end{equation}

The heat current operator can be conveniently written as a sum of two contributions
\begin{equation}
\widehat{\jmath
}_{ij}^{(h)} = \widehat{\jmath }_{ij}^{(h,0)} +  \widehat{\jmath
}_{ij}^{(h,1)},
\end{equation}
where, the non-interacting part $\widehat{\jmath }_{ij}^{(h,0)}$ of the heat
current originates from the sum of the commutators \bigskip\ $\widehat{\mathcal{C}}_{i}^{(et)}+$ $
\widehat{\mathcal{C}}_{i}^{(te)}$ and the interacting part $\widehat{\jmath
}_{ij}^{(h,1)}$ from the sum of $\widehat{\mathcal{C}}_{i}^{(ct)}+$ $\widehat{\mathcal{
C}}_{i}^{(tc)}$.

\subsubsection{non-interacting part of heat current operator}

We first calculate the non-interacting part of the heat current operator, $\widehat{\jmath }_{ij}^{(h,0)}$.
That is, we consider the sum of commutators \bigskip\ $\widehat{\mathcal{C}}_{i}^{(et)}+$ $
\widehat{\mathcal{C}}_{i}^{(te)}$. A straightforward calculation leads to
\begin{widetext}
\begin{eqnarray}
\widehat{\mathcal{C}}_{i}^{(et)} &=&\frac{1}{2}\underset{i^{\prime
},j,k^{\prime },q,k}{\sum }\xi _{k}\left( \widehat{a}_{i,k}^{\dagger }
\widehat{a}_{i,k}\left[ t_{i^{\prime }j}^{k^{\prime }q}\widehat{a}
_{i^{\prime },k^{\prime }}^{\dagger }\widehat{a}_{j,q}+t_{ji^{\prime
}}^{qk^{\prime }}\widehat{a}_{j,q}^{\dagger }\widehat{a}_{i^{\prime
},k^{\prime }}\right] -\left[ t_{i^{\prime }j}^{k^{\prime }q}\widehat{a}
_{i^{\prime },k^{\prime }}^{\dagger }\widehat{a}_{j,q}+t_{ji^{\prime
}}^{qk^{\prime }}\widehat{a}_{j,q}^{\dagger }\widehat{a}_{i^{\prime
},k^{\prime }}\right] \widehat{a}_{i,k}^{\dagger }\widehat{a}_{i,k}\right)\nn \\
&=&\underset{j,k,q}{\sum }\xi _{k}\left( t_{ij}^{kq}\widehat{a}
_{i,k}^{\dagger }\widehat{a}_{j,q}-t_{ji}^{qk}\widehat{a}_{j,q}^{\dagger }
\widehat{a}_{i,k}\right) , \label{Cet}\\
\widehat{\mathcal{C}}_{i}^{(te)} &=&\frac{1}{2}\underset{j,k,q,i^{\prime
},k^{\prime }}{\sum }\xi _{k^{\prime }}\left( \left[ t_{ij}^{kq}\widehat{a}%
_{i,k}^{\dagger }\widehat{a}_{j,q}+t_{ji}^{qk}\widehat{a}_{j,q}^{\dagger }%
\widehat{a}_{i,k}\right] \widehat{a}_{i^{\prime },k^{\prime }}^{\dagger }%
\widehat{a}_{i^{\prime },k^{\prime }}-\widehat{a}_{i^{\prime },k^{\prime
}}^{\dagger }\widehat{a}_{i^{\prime },k^{\prime }}\left[ t_{ij}^{kq}\widehat{%
a}_{i,k}^{\dagger }\widehat{a}_{j,q}+t_{ji}^{qk}\widehat{a}_{j,q}^{\dagger }%
\widehat{a}_{i,k}\right] \right)\nn \\
&=&\frac{1}{2}\underset{j,k,q}{\sum }\left( \xi _{q}t_{ij}^{kq}\widehat{a}%
_{i,k}^{\dagger }\widehat{a}_{j,q}+\xi _{k}t_{ji}^{qk}\widehat{a}%
_{j,q}^{\dagger }\widehat{a}_{i,k}-\xi _{k}t_{ij}^{kq}\widehat{a}%
_{i,k}^{\dagger }\widehat{a}_{j,q}-\xi _{q}t_{ji}^{qk}\widehat{a}%
_{j,q}^{\dagger }\widehat{a}_{i,k}\right). \label{Cte}
\end{eqnarray}
\end{widetext}
Using Eqs.~(\ref{Cet}) and (\ref{Cte}) we obtain for the sum
\begin{eqnarray}
\widehat{\mathcal{C}}_{i}^{(et)}+\widehat{\mathcal{C}}
_{i}^{(te)} &=&\frac{1}{2}\underset{j,k,q}{\sum }\left( \left( \xi _{q}+\xi _{k}\right)
t_{ij}^{kq}\widehat{a}_{i,k}^{\dagger }\widehat{a}_{j,q}  \right. \nonumber \\
&&  \left. -\left( \xi _{k} + \xi
_{q}\right) t_{ji}^{qk}\widehat{a}_{j,q}^{\dagger }\widehat{a}_{i,k} \right).
\end{eqnarray}
As a result the non-interacting part of the heat current operator of granular metals has the form
\begin{equation}\label{jh0}
\widehat{\jmath }_{ij}^{(h,0)}=\imath \underset{k,q}{\sum }%
\frac{\xi _{k}+\xi _{q}}{2}\left[ t_{ij}^{kq}\widehat{a}_{i,k}^{\dagger }%
\widehat{a}_{j,q}-t_{ji}^{qk}\widehat{a}_{j,q}^{\dagger }\widehat{a}_{i,k}%
\right]\,.
\end{equation}

\subsubsection{interacting part of heat current operator}

To obtain the expression for the interacting part of the heat current operator
$\widehat{\jmath }_{ij}^{(h,1)}$ we need to consider the commutators of the form $\left[
\widehat{\epsilon }_{i^{\prime }j}^{(t)};\widehat{n}_{i}\right] $, where
 $\widehat{\epsilon }_{ij}^{(t)}=\frac{1}{2}\underset{k,q}{\sum }
\left[ t_{ij}^{kq}\widehat{a}_{i,k}^{\dagger }\widehat{a}_{j,q}+t_{ji}^{qk}
\widehat{a}_{j,q}^{\dagger }\widehat{a}_{i,k}\right]$ .
Using the calculation for $\widehat{\mathcal{C}}_{i}^{(nt)}$ [Eq.~(\ref{eq.Cnt})] we get
\begin{widetext}
\begin{eqnarray}
2\left[ \widehat{\epsilon }_{i^{\prime }j}^{(t)};\widehat{n}_{i}\right] &=&%
\underset{k,k^{\prime },q}{\sum }\left( t_{i^{\prime }j}^{k^{\prime
}q}\left( \delta _{ij}\delta _{kq}\widehat{a}_{i^{\prime },k^{\prime
}}^{\dagger }\widehat{a}_{i,k}-\delta _{ii^{\prime }}\delta _{kk^{\prime }}%
\widehat{a}_{i,k}^{\dagger }\widehat{a}_{j,q}\right) +t_{ji^{\prime
}}^{qk^{\prime }}\left( \delta _{ii^{\prime }}\delta _{kk^{\prime }}\widehat{%
a}_{j,q}^{\dagger }\widehat{a}_{i,k}-\delta _{ij}\delta _{kq}\widehat{a}%
_{i,k}^{\dagger }\widehat{a}_{i^{\prime },k^{\prime }}\right) \right) \\
&=&\left[ \delta _{ij}-\delta _{ii^{\prime }}\right] \underset{k,q}{\sum }%
\left( t_{i^{\prime }j}^{kq}\widehat{a}_{i^{\prime },k}^{\dagger }\widehat{a}%
_{j,q}-t_{ji^{\prime }}^{kq}\widehat{a}_{j,k}^{\dagger }\widehat{a}%
_{i^{\prime },q}\right) =-\frac{\imath }{e}\left[ \delta _{ij}-\delta
_{ii^{\prime }}\right] \widehat{\jmath }_{i^{\prime }j}^{(e)}. \nonumber
\end{eqnarray}
For the following steps of the calculation of $\widehat{\mathcal{C}}_{i}^{(ct)}$ and $\widehat{\mathcal{C}}_{i}^{(tc)}$, we need the commutator $\left[ \widehat{n}_{m};\widehat{\jmath }_{ij}^{(e)}\right]$
\begin{eqnarray}
-\frac{\imath }{e}\left[ \widehat{n}_{m};\widehat{\jmath }_{ij}^{(e)}\right]
&=&\underset{k,q}{\sum }\left[ \widehat{n}_{m}\left( t_{ij}^{kq}\widehat{a}%
_{i,k}^{\dagger }\widehat{a}_{j,q}-t_{ji}^{qk}\widehat{a}_{j,q}^{\dagger }%
\widehat{a}_{i,k}\right) -\left( t_{ij}^{kq}\widehat{a}_{i,k}^{\dagger }%
\widehat{a}_{j,q}-t_{ji}^{qk}\widehat{a}_{j,q}^{\dagger }\widehat{a}%
_{i,k}\right) \widehat{n}_{m}\right] \nonumber\\
&=&-\left( \delta _{mj}-\delta _{mi}\right) \underset{k,q}{\sum }\left(
t_{ij}^{kq}\widehat{a}_{i,k}^{\dagger }\widehat{a}_{j,q}+t_{ji}^{qk}\widehat{%
a}_{j,q}^{\dagger }\widehat{a}_{i,k}\right) =2\left( \delta _{mi}-\delta
_{mj}\right) \widehat{\epsilon }_{ij}^{(t)}. \label{eq.comNje}
\end{eqnarray}
To calculate the interacting part of the heat current operator $\widehat{\jmath }_{ij}^{(h,1)}$ we need
the following commutators
\begin{eqnarray}
\frac{4}{e^{2}}\widehat{\mathcal{C}}_{i}^{(ct)} &=&\underset{j,i^{\prime
},j^{\prime }}{2\sum }\left( \widehat{n}_{i}C_{ij}^{-1}\widehat{n}_{j}%
\widehat{\epsilon }_{i^{\prime }j^{\prime }}^{(t)}-\widehat{\epsilon }%
_{i^{\prime }j^{\prime }}^{(t)}\widehat{n}_{i}C_{ij}^{-1}\widehat{n}%
_{j}\right) = \frac{\imath }{e}\underset{j,i^{\prime },j^{\prime }}{\sum }%
C_{ij}^{-1}\left( \left[ \delta _{ij^{\prime }}-\delta _{ii^{\prime }}\right]
\widehat{\jmath }_{i^{\prime }j^{\prime }}^{(e)}\widehat{n}_{j}+\left[
\delta _{jj^{\prime }}-\delta _{ji^{\prime }}\right] \widehat{n}_{i}\widehat{%
\jmath }_{i^{\prime }j^{\prime }}^{(e)}\right) \\
&=&\frac{\imath }{e}\underset{j}{\sum }C_{ij}^{-1}\left[ \underset{i^{\prime
}}{\sum }\widehat{\jmath }_{i^{\prime }i}^{(e)}\widehat{n}_{j}-\underset{%
j^{\prime }}{\sum }\widehat{\jmath }_{ij^{\prime }}^{(e)}\widehat{n}_{j}+%
\underset{i^{\prime }}{\sum }\widehat{n}_{i}\widehat{\jmath }_{i^{\prime
}j}^{(e)}-\underset{j^{\prime }}{\sum }\widehat{n}_{i}\widehat{\jmath }%
_{jj^{\prime }}^{(e)}\right] \nonumber \\
&=&\frac{\imath }{e}\underset{j,m}{\sum }C_{ij}^{-1}\left[ \left( \widehat{%
\jmath }_{mi}^{(e)}-\widehat{\jmath }_{im}^{(e)}\right) \widehat{n}_{j}+%
\widehat{n}_{i}\left( \widehat{\jmath }_{mj}^{(e)}-\widehat{\jmath }%
_{jm}^{(e)}\right) \right], \nonumber\\
\frac{4}{e^{2}}\widehat{\mathcal{C}}_{i}^{(tc)} &=&\underset{j,i^{\prime
},j^{\prime }}{2\sum }\left( \widehat{\epsilon }_{ij}^{(t)}\widehat{n}%
_{i^{\prime }}C_{i^{\prime }j^{\prime }}^{-1}\widehat{n}_{j^{\prime }}-%
\widehat{n}_{i^{\prime }}C_{i^{\prime }j^{\prime }}^{-1}\widehat{n}%
_{j^{\prime }}\widehat{\epsilon }_{ij}^{(t)}\right) =
 -\frac{\imath }{e}\underset{j,i^{\prime },j^{\prime }}{\sum }C_{i^{\prime
}j^{\prime }}^{-1}\left( \left[ \delta _{j^{\prime }j}-\delta _{j^{\prime }i}%
\right] \widehat{n}_{i^{\prime }}\widehat{\jmath }_{ij}^{(e)}+\left[ \delta
_{i^{\prime }j}-\delta _{i^{\prime }i}\right] \widehat{\jmath }_{ij}^{(e)}%
\widehat{n}_{j^{\prime }}\right) \\
&=&-\frac{\imath }{e}\underset{j}{\sum }\left[ \underset{i^{\prime }}{\sum }%
C_{i^{\prime }j}^{-1}\widehat{n}_{i^{\prime }}\widehat{\jmath }_{ij}^{(e)}+%
\underset{j^{\prime }}{\sum }C_{jj^{\prime }}^{-1}\widehat{\jmath }%
_{ij}^{(e)}\widehat{n}_{j^{\prime }}-\underset{i^{\prime }}{\sum }%
C_{i^{\prime }i}^{-1}\widehat{n}_{i^{\prime }}\widehat{\jmath }_{ij}^{(e)}-%
\underset{j^{\prime }}{\sum }C_{ij^{\prime }}^{-1}\widehat{\jmath }%
_{ij}^{(e)}\widehat{n}_{j^{\prime }}\right] \nonumber \\
&=&-\frac{\imath }{e}\underset{j,m}{\sum }\left[ \left(
C_{mj}^{-1}-C_{mi}^{-1}\right) \widehat{n}_{m}\widehat{\jmath }%
_{ij}^{(e)}+\left( C_{jm}^{-1}-C_{im}^{-1}\right) \widehat{\jmath }%
_{ij}^{(e)}\widehat{n}_{m}\right]. \nonumber
\end{eqnarray}
Using the symmetry relations: $\widehat{\jmath }_{mi}^{(e)}=-\widehat{\jmath }%
_{im}^{(e)}$ and $C_{mj}^{-1}=C_{jm}^{-1}$, and $\left[ \widehat{n}_{m};%
\widehat{\jmath }_{ij}^{(e)}\right] =2\imath e\left( \delta _{mi}-\delta
_{mj}\right) \widehat{\epsilon }_{ij}^{(t)}$ [Eq.~\ref{eq.comNje}] we obtain
\begin{eqnarray}
\widehat{\mathcal{C}}_{i}^{(ct)}+\widehat{\mathcal{C}}%
_{i}^{(tc)} &=&\frac{\imath e}{4}\underset{j,m}{\sum }\left( C_{ij}^{-1}%
\left[ \left( \widehat{\jmath }_{mi}^{(e)}-\widehat{\jmath }%
_{im}^{(e)}\right) \widehat{n}_{j}+\widehat{n}_{i}\left( \widehat{\jmath }%
_{mj}^{(e)}-\widehat{\jmath }_{jm}^{(e)}\right) \right] -\left[ \left(
C_{mj}^{-1}-C_{mi}^{-1}\right) \widehat{n}_{m}\widehat{\jmath }%
_{ij}^{(e)}+\left( C_{jm}^{-1}-C_{im}^{-1}\right) \widehat{\jmath }%
_{ij}^{(e)}\widehat{n}_{m}\right] \right)  \notag \\
&=&\frac{\imath e}{4}\underset{j,m}{\sum }\left( 2C_{ij}^{-1}\left[ \widehat{%
\jmath }_{mi}^{(e)}\widehat{n}_{j}+\widehat{n}_{i}\widehat{\jmath }%
_{mj}^{(e)}\right] -\left[ C_{mj}^{-1}\left\{ \widehat{n}_{m};\widehat{%
\jmath }_{ij}^{(e)}\right\} _{+}-C_{mi}^{-1}\left\{ \widehat{\jmath }%
_{ij}^{(e)};\widehat{n}_{m}\right\} _{+}\right] \right)  \notag \\
&=&\frac{\imath e}{4}\underset{j,m}{\sum }C_{ij}^{-1}\left[ \widehat{\jmath }%
_{mi}^{(e)}\widehat{n}_{j}+\widehat{n}_{j}\widehat{\jmath }%
_{mi}^{(e)}-2\imath e\left( \delta _{jm}-\delta _{ji}\right) \widehat{%
\epsilon }_{mi}^{(t)}+\widehat{n}_{i}\widehat{\jmath }_{mj}^{(e)}+\widehat{%
\jmath }_{mj}^{(e)}\widehat{n}_{i}+2\imath e\left( \delta _{im}-\delta
_{ij}\right) \widehat{\epsilon }_{mj}^{(t)}\right]  \notag \\
&&-\left[ C_{mj}^{-1}\left\{ \widehat{n}_{m};\widehat{\jmath }%
_{ij}^{(e)}\right\} _{+}-C_{mi}^{-1}\left\{ \widehat{\jmath }_{ij}^{(e)};%
\widehat{n}_{m}\right\} _{+}\right] \nonumber \\
&=&\frac{-e^{2}}{2}\underset{j}{\sum }\underset{=0}{\underbrace{\left( -%
\widehat{\epsilon }_{ji}^{(t)}+\widehat{\epsilon }_{ji}^{(t)}+\widehat{%
\epsilon }_{ij}^{(t)}-\widehat{\epsilon }_{ji}^{(t)}\right) }} \nonumber \\
&&+\frac{\imath e}{4}\underset{j,m}{\sum }C_{ij}^{-1}\left[ %
\left\{ \widehat{\jmath }_{mi}^{(e)};\widehat{n}_{j}\right\} _{+}+\left\{
\widehat{n}_{i};\widehat{\jmath }_{mj}^{(e)}\right\} _{+}\right] -\left[
C_{mj}^{-1}\left\{ \widehat{n}_{m};\widehat{\jmath }_{ij}^{(e)}\right\} _{+}-%
C_{ji}^{-1}\left\{ \widehat{\jmath }_{im}^{(e)};\widehat{n}%
_{j}\right\} _{+}\right] \nonumber \\
&=&\frac{\imath e}{4}\underset{j,m}{\sum }\left( C_{ij}^{-1}\left\{ \widehat{%
n}_{i};\widehat{\jmath }_{mj}^{(e)}\right\} _{+}-C_{mj}^{-1}\left\{ \widehat{%
n}_{m};\widehat{\jmath }_{ij}^{(e)}\right\} _{+}\right).
\end{eqnarray}
As a result for the interaction part of the heat current operator of granular metals we obtain
\begin{equation}\label{jh1}
\widehat{\jmath }_{ij}^{(h,1)}=-\frac{e}{4}\underset{m}{
\sum }\left( C_{im}^{-1}\left\{ \widehat{n}_{i};\widehat{\jmath }
_{jm}^{(e)}\right\} _{+}-C_{jm}^{-1}\left\{ \widehat{n}_{j};\widehat{\jmath }
_{im}^{(e)}\right\} _{+}\right)\,.
\end{equation}

So far we have omitted the last commutator $\widehat{\mathcal{C}}_{i}^{(tt)}$ in Eq.~(\ref{b7}), which would be an additional contribution to the non-interacting part of the heat current operator, $\widehat{\jmath }_{ij}^{(h,0)}$ in Eq.~(\ref{jh0}).
However, if this term is summed over $i$ it vanishes and does not contribute to $\widehat{\jmath }_{ij}^{(h,0)}$. However,
for completeness we present the calculation of $\widehat{\mathcal{C}}_{i}^{(tt)}$ here as well.

We define the commutator $\widehat{\mathcal{C}}_{i}^{(tt)}$ as
\begin{equation}
\widehat{\mathcal{C}}_{i}^{(tt)}=\underset{j,i^{\prime },j^{\prime }}{%
\sum }\left( \widehat{\epsilon }_{ij}^{(t)}\widehat{\epsilon }_{i^{\prime
}j^{\prime }}^{(t)}-\widehat{\epsilon }_{i^{\prime }j^{\prime }}^{(t)}
\widehat{\epsilon }_{ij}^{(t)}\right)\equiv \widehat{A}-\widehat{B},
\end{equation}
and simplify first the expression for operator $4\widehat{B}$.
\begin{eqnarray}
4\widehat{\epsilon }_{i^{\prime }j^{\prime }}^{(t)}\widehat{\epsilon }%
_{ij}^{(t)} &=&\underset{k,q,k^{\prime },q^{\prime }}{\sum }\left(
t_{i^{\prime }j^{\prime }}^{k^{\prime }q^{\prime }}\widehat{a}_{i^{\prime
},k^{\prime }}^{\dagger }\widehat{a}_{j^{\prime },q^{\prime }}+t_{j^{\prime
}i^{\prime }}^{q^{\prime }k^{\prime }}\widehat{a}_{j^{\prime },q^{\prime
}}^{\dagger }\widehat{a}_{i^{\prime },k^{\prime }}\right) \left( t_{ij}^{kq}%
\widehat{a}_{i,k}^{\dagger }\widehat{a}_{j,q}+t_{ji}^{qk}\widehat{a}%
_{j,q}^{\dagger }\widehat{a}_{i,k}\right) \\
&=&\underset{k,q,k^{\prime },q^{\prime }}{\sum }\left( t_{i^{\prime
}j^{\prime }}^{k^{\prime }q^{\prime }}t_{ij}^{kq}\left[ \delta _{j^{\prime
}i}\delta _{q^{\prime }k}\widehat{a}_{i^{\prime },k^{\prime }}^{\dagger }%
\widehat{a}_{j,q}-\delta _{i^{\prime }j}\delta _{k^{\prime }q}\widehat{a}%
_{i,k}^{\dagger }\widehat{a}_{j^{\prime },q^{\prime }}\right] +t_{j^{\prime
}i^{\prime }}^{q^{\prime }k^{\prime }}t_{ij}^{kq}\left[ \delta _{i^{\prime
}i}\delta _{k^{\prime }k}\widehat{a}_{j^{\prime },q^{\prime }}^{\dagger }%
\widehat{a}_{j,q}-\delta _{j^{\prime }j}\delta _{q^{\prime }q}\widehat{a}%
_{i,k}^{\dagger }\widehat{a}_{i^{\prime },k^{\prime }}\right] \right. + \nonumber \\
&&+\left. t_{i^{\prime }j^{\prime }}^{k^{\prime }q^{\prime }}t_{ji}^{qk}
\left[ \delta _{j^{\prime }j}\delta _{q^{\prime }q}\widehat{a}_{i^{\prime
},k^{\prime }}^{\dagger }\widehat{a}_{i,k}-\delta _{i^{\prime }i}\delta
_{k^{\prime }k}\widehat{a}_{j,q}^{\dagger }\widehat{a}_{j^{\prime
},q^{\prime }}\right] +t_{j^{\prime }i^{\prime }}^{q^{\prime }k^{\prime
}}t_{ji}^{qk}\left[ \delta _{i^{\prime }j}\delta _{k^{\prime }q}\widehat{a}%
_{j^{\prime },q^{\prime }}^{\dagger }\widehat{a}_{i,k}-\delta _{j^{\prime
}i}\delta _{q^{\prime }k}\widehat{a}_{j,q}^{\dagger }\widehat{a}_{i^{\prime
},k^{\prime }}\right] \right) +4\widehat{A} \,.\nonumber
\end{eqnarray}
And the final step is:
\begin{eqnarray}
\widehat{\mathcal{C}}_{i}^{(tt)} &=&-\frac{1}{4}\underset{j,k,q}{\sum }\left[
\underset{i^{\prime },k^{\prime }}{\sum }\left( t_{i^{\prime }i}^{k^{\prime
}k}t_{ij}^{kq}\widehat{a}_{i^{\prime },k^{\prime }}^{\dagger }\widehat{a}%
_{j,q}-t_{ji^{\prime }}^{qk^{\prime }}t_{ij}^{kq}\widehat{a}_{i,k}^{\dagger }%
\widehat{a}_{i^{\prime },k^{\prime }}\right) -\underset{j^{\prime
},q^{\prime }}{\sum }\left( t_{jj^{\prime }}^{qq^{\prime }}t_{ij}^{kq}%
\widehat{a}_{i,k}^{\dagger }\widehat{a}_{j^{\prime },q^{\prime
}}-t_{j^{\prime }i}^{q^{\prime }k}t_{ij}^{kq}\widehat{a}_{j^{\prime
},q^{\prime }}^{\dagger }\widehat{a}_{j,q}\right) \right. \\
&&+\left. \underset{i^{\prime },k^{\prime }}{\sum }\left( t_{i^{\prime
}j}^{k^{\prime }q}t_{ji}^{qk}\widehat{a}_{i^{\prime },k^{\prime }}^{\dagger }
\widehat{a}_{i,k}-t_{ii^{\prime }}^{kk^{\prime }}t_{ji}^{qk}\widehat{a}
_{j,q}^{\dagger }\widehat{a}_{i^{\prime },k^{\prime }}\right) -\underset{
j^{\prime },q^{\prime }}{\sum }\left( t_{ij^{\prime }}^{kq^{\prime}}t_{ji}^{qk}\widehat{a}_{j,q}^{\dagger }\widehat{a}_{j^{\prime },q^{\prime}}-t_{j^{\prime }j}^{q^{\prime }q}t_{ji}^{qk}\widehat{a}_{j^{\prime},q^{\prime }}^{\dagger }\widehat{a}_{i,k}\right) \right] \nonumber \\
&=&-\frac{1}{4}\underset{j,m,k,q,p}{\sum }\left[ t_{mi}^{pk}t_{ij}^{kq}%
\widehat{a}_{m,p}^{\dagger }\widehat{a}_{j,q}+t_{mi}^{pk}t_{ij}^{kq}\widehat{%
a}_{m,p}^{\dagger }\widehat{a}_{j,q}-t_{jm}^{qp}t_{ij}^{kq}\widehat{a}%
_{i,k}^{\dagger }\widehat{a}_{m,p}-t_{jm}^{qp}t_{ij}^{kq}\widehat{a}%
_{i,k}^{\dagger }\widehat{a}_{m,p}\right.\nn \\
&&+\left. t_{mj}^{pq}t_{ji}^{qk}\widehat{a}_{m,p}^{\dagger }\widehat{a}%
_{i,k}+t_{mj}^{pq}t_{ji}^{qk}\widehat{a}_{m,p}^{\dagger }\widehat{a}%
_{i,k}-t_{im}^{kp}t_{ji}^{qk}\widehat{a}_{j,q}^{\dagger }\widehat{a}%
_{m,p}-t_{im}^{kp}t_{ji}^{qk}\widehat{a}_{j,q}^{\dagger }\widehat{a}_{m,p}%
\right]\nn \\
&=&-\frac{1}{2}\underset{j,m,k,q,p}{\sum }\left[ \underline{%
t_{mi}^{pk}t_{ij}^{kq}\widehat{a}_{m,p}^{\dagger }\widehat{a}_{j,q}}%
-t_{jm}^{qp}t_{ij}^{kq}\widehat{a}_{i,k}^{\dagger }\widehat{a}%
_{m,p}+t_{mj}^{pq}t_{ji}^{qk}\widehat{a}_{m,p}^{\dagger }\widehat{a}_{i,k}-%
\underline{t_{im}^{kp}t_{ji}^{qk}\widehat{a}_{j,q}^{\dagger }\widehat{a}%
_{m,p}}\right] \nonumber \\
&=&\frac{1}{2}\underset{j,m,k,q,p}{\sum }\left[ t_{jm}^{qp}t_{ij}^{kq}%
\widehat{a}_{i,k}^{\dagger }\widehat{a}_{m,p}-t_{mj}^{pq}t_{ji}^{qk}\widehat{%
a}_{m,p}^{\dagger }\widehat{a}_{i,k}\right] =\underset{j,m,k,q,p}{\sum }%
\frac{t_{jm}^{qp}t_{ij}^{kq}}{2}\left[ \widehat{a}_{i,k}^{\dagger }\widehat{a%
}_{m,p}-\widehat{a}_{m,p}^{\dagger }\widehat{a}_{i,k}\right]\,. \nonumber
\end{eqnarray}
\end{widetext}
The underlined terms cancel since $t_{mi}^{pk}=t_{im}^{kp}$, which is also
used in the last step. If this expression is summed over $i$, we can
exchange indices $m$ and $i$ (and $k$ and $q$) in the second summand.
As a result we obtain
\begin{equation}
\underset{i}{\sum }\widehat{\mathcal{C}}_{i}^{(tt)}=0,
\end{equation}
i.e. there is no additional contribution to the full heat current operator $\overrightarrow{\jmath }_{h}$ in Eq.~(\ref{b4}) from this term.

\subsection{summary: electric and heat current operators}

To summarize this appendix we explicitly write the expressions for electric $\widehat{\jmath }_{ij}^{(e)}$ and heat $\hat{\jmath }_{ij}^{(h)}$ current operators
\begin{equation}
\widehat{\jmath }_{ij}^{(e)}=\imath e\underset{k,q}{\sum }
\left( t_{ij}^{kq}\widehat{a}_{i,k}^{\dagger }\widehat{a}_{j,q}-t_{ji}^{qk}
\widehat{a}_{j,q}^{\dagger }\widehat{a}_{i,k}\right)\,.
\end{equation}
\begin{eqnarray}
\hat{\jmath }_{ij}^{(h)} &=& \hat{\jmath }_{ij}^{(h,0)} + \hat{\jmath }_{ij}^{(h,1)}, \\
\hat{\jmath }_{ij}^{(h,0)}&=&\imath \underset{k,q}{\sum }
\frac{\xi _{k}+\xi _{q}}{2}\left[ t_{ij}^{kq} \hat{a}_{ik}^{\dagger }
\hat{a}_{jq}-t_{ji}^{qk} \hat{a}_{jq}^{\dagger } \hat{a}_{ik} \nonumber
\right],\\ \nonumber
\hat{\jmath }_{ij}^{(h,1)}&=&-\frac{e}{4}\underset{m}{
\sum } \left[ \frac{\{ \hat{n}_{i};\hat{\jmath }
_{jm}^{(e)} \} _{+}}{C_{im}}-\frac{ \{ \hat{n}_{j};\hat{\jmath }
_{im}^{(e)} \} _{+}}{C_{jm}} \right].
\end{eqnarray}

\section{Thermoelectric coefficient of granular metals in the absence of interaction}
\label{app.eta0}

\begin{figure}[t]
\includegraphics[width=0.8\linewidth]{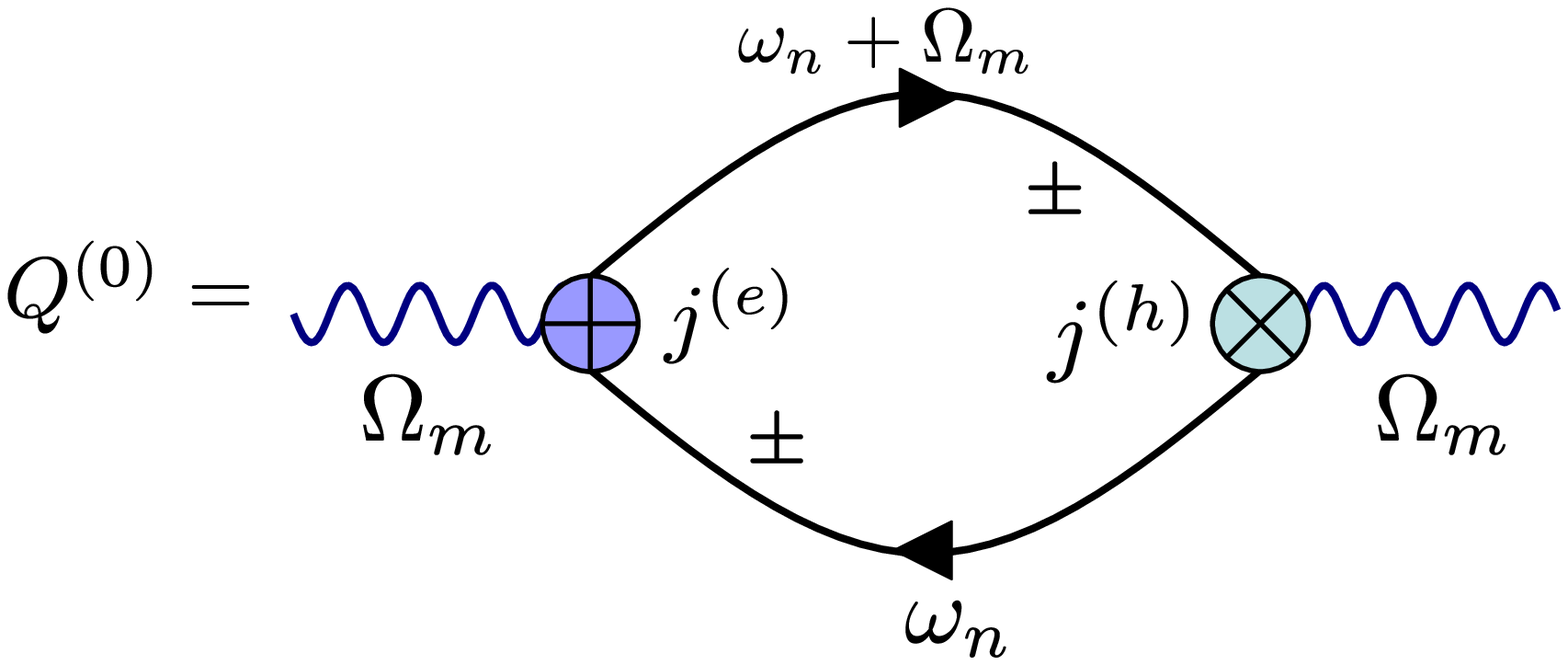}
\caption{(color online) Lowest order diagram for the heat-electric current correlator of granular metals.
The external Bosonic frequency is denoted by $\Omega$ (wavy lines) and the internal Fermionic frequency by $\omega$. The electric and heat current vertexes are $\protect\overrightarrow{\jmath_e}$ and $\protect\overrightarrow{\jmath_h}$, respectively. The $\pm$ denote the possible analytical structure of the Green's functions (straight lines).}\label{fig.Q0}
\end{figure}

In this appendix we consider the thermoelectric coefficient $\eta^{(0)}$ of granular metals in the absence of interaction in analogy to Appendix~\ref{app.hom}. The expression for the thermoelectric coefficient in linear response theory is
\begin{equation}
\label{eta0grain}
\eta ^{(0)}=\imath \left. \frac{\partial }{a^{d}T\partial \Omega }%
\right\vert _{\Omega =0}Q^{(0)}.
\end{equation}
Here $a$ is the grain size and $Q^{(0)}$ the correlator of the heat current, $\overrightarrow{\jmath }_{h}^{(0)}$ (see also Fig.~\ref{fig.jver}a), and electric current, $\overrightarrow{\jmath }_{e}$, shown in Fig.~\ref{fig.Q0}.

For granular metals the important element of the diagram is the tunneling matrix
elements $t_{ij}^{kq}$ describing the coupling between grains $i$ and $j$. Therefore we derive an expression for $t_{ij}^{kq}$ in the following,
assuming that i and j are nearest neighbor grains and $t_{ij}^{kq}$ is independent
of the position in the sample. In order to calculate the energy dependence of these elements we
assume the tunneling barrier between grains to be a delta potential.
For the one-particle Hamiltonian $\widehat{H}=-\frac{\hbar ^{2}}{2m}\frac{d^{2}}{
dx^{2}}+\lambda \delta \left( x\right) $ the transmission rate for a single
particle with energy $\varepsilon _{p}=\varepsilon _{F}+\xi _{p}$ is
\begin{eqnarray}
T_{p} = \left( 1+\frac{m\lambda ^{2}}{2\hbar ^{2}\varepsilon _{p}}\right)
^{-1}\simeq \left( 1+\frac{m\lambda ^{2}}{2\hbar ^{2}\varepsilon _{F}}\left(
1-\xi _{p}/\varepsilon _{F}\right) \right) ^{-1} \\
 = T_{0}\left( 1+\frac{2\hbar
^{2}\varepsilon _{F}}{m\lambda ^{2}}-\xi _{p}/\varepsilon _{F}\right) ^{-1} \nonumber\,.
\end{eqnarray}
Here $T_0 \left( 1+\frac{m\lambda ^{2}}{2\hbar ^{2}\varepsilon _F}\right)^{-1}$ is the transmission rate at $\varepsilon _{p}=\varepsilon _{F}$
and we use the fact that $\xi_p \ll \varepsilon _{F}$.
Next, we consider the case of large barriers, in this regime $T_{p}\simeq T_{0}\left( 1+\xi
_{p}/\varepsilon _{F}\right) $. In granular systems we have many channels
and have to consider tunneling processes with energy $\xi _{1}$ in grain $i=1$
and with $\xi _{2}$ in grain $j=2$: $t^{2}\propto N\left(
T_{p_{1}}^{2}+T_{p_{2}}^{2}\right) $.
So, we obtain
\begin{equation}
t^{2}(\xi _{1},\xi _{2})\simeq t_{0}^{2}\left( 1+\frac{\xi _{1}+\xi _{2}}{%
\varepsilon _{F}}\right)\,.
\end{equation}
Therefore we have the following expression for correlation function in Eq.~(\ref{eta0grain})
\begin{widetext}
\begin{eqnarray}
\label{Q0}
Q^{(0)} &=&-set_{0}^{2}T\left\langle \overrightarrow{n}_{e}^{(0)}\cdot
\overrightarrow{n}_{h}^{(0)}\right\rangle \underset{\omega _{n}}{\sum }
a^{2d+2}\int \frac{d^{d}p_{1}}{(2\pi )^{d}}\int \frac{d^{d}p_{2}}{(2\pi
)^{d}}\left( \frac{\xi _{1}+\xi _{2}}{2}\right) \left( 1+\frac{\xi _{1}+\xi
_{2}}{\varepsilon _{F}}\right) G(p_{1},\omega _{n})G(p_{2},\omega
_{n}+\Omega _{m}) \nonumber \\
&\approx &-\frac{s}{2d}et_{0}^{2}Ta^{2d+2}\left( \nu _{d}^{(0)}\right) ^{2}%
\underset{\omega _{n}}{\sum }\int d\xi _{1}d\xi _{2}\ g(\xi _{1},\xi
_{2})G(\xi _{1},\omega _{n})G(\xi _{2},\omega _{n}+\Omega _{m})\,,
\end{eqnarray}
\end{widetext}
where $\overrightarrow{n}^{(0)}_{\alpha}$ is the unit vector in direction of the current $\alpha\in\{e,h\}$, $\langle \overrightarrow{n}_{e}^{(0)}\cdot
\overrightarrow{n}_{h}^{(0)}\rangle = 1/d$ is the result of averaging over angles, the summation goes over Fermionic Matsubara
frequencies $\omega_n = 2\pi T (n + 1/2)$, and
$G(p,\omega_n)$ is the Green's function defined in Eq.~(\ref{GF}) of Appendix~\ref{app.hom} with momenta/energies $p_i$/$\xi_i$ of grain $i$. To shorter the notation in the following we neglect the momentum argument and attach the grain index to $G$.
In Eq.~(\ref{Q0}) we introduce the notation
\begin{eqnarray}
g(\xi _{1},\xi _{2})=\left( \xi _{1}+\xi _{2}\right) \left( 1+\frac{
\xi _{1}+\xi _{2}}{\varepsilon _{F}}\right) \\
\times \left( 1+\left[ \frac{d}{2}-1
\right] \frac{\xi _{1}}{\varepsilon _{F}}\right) \left( 1+\left[ \frac{d}{2}
-1\right] \frac{\xi _{2}}{\varepsilon _{F}}\right). \nonumber
\end{eqnarray}
The factors in this order arise from: the heat current vertex Eq.~(\ref{12c}), the energy correction to tunneling elements Eq.~(\ref{t}), and corrections to the DOS due to finite Fermi energy Eq.~(\ref{transfomration}). In the linear order in $\xi /\varepsilon _{F}$ we obtain
\begin{equation}
\label{g12}
g(\xi _{1},\xi _{2}) = \xi _{1}+\xi _{2}+\frac{d}{2}\frac{\left( \xi
_{1}+\xi _{2}\right) ^{2}}{\varepsilon _{F}}.
\end{equation}

We first perform the analytical continuation in Eq.~(\ref{Q0}) (for convenience the grain index is
written as index to the Green's functions)
\begin{widetext}
\begin{equation}
\label{sum}
\int d\xi _{1}d\xi _{2}\ g(\xi _{1},\xi _{2})\left[ \underset{S_{1}}{%
\underbrace{\underset{n\in I_{1}}{\sum }G_{1}^{-}(\omega
_{n})G_{2}^{-}(\omega _{n}+\Omega _{m})}}+\underset{S_{2}}{\underbrace{%
\underset{n\in I_{2}}{\sum }G_{1}^{+}(\omega _{n})G_{2}^{-}(\omega
_{n}+\Omega _{m})}}+\underset{S_{3}}{\underbrace{\underset{n\in I_{3}}{\sum }%
G_{1}^{+}(\omega _{n})G_{2}^{+}(\omega _{n}+\Omega _{m})}}\right].
\end{equation}
After analytical continuation we obtain
\begin{eqnarray}
S_{1} &=& - \int \frac{d\omega}{4\pi \imath T} \tanh \left( \frac{\omega}{2T}
\right) G_{1}^{-}(-\imath \omega +\imath \Omega )G_{2}^{-}(-\imath
\omega ), \\
S_{2} &=& - \int \frac{d\omega}{4\pi \imath T} \tanh \left( \frac{\omega}{2T}
\right) \left[ G_{1}^{-}(-\imath \omega )G_{2}^{+}(-\imath \omega
-\imath \Omega ) -G_{1}^{-}(-\imath \omega +\imath \Omega )G_{2}^{+}(-\imath
\omega )\right], \\
S_{3} &=& - \int \frac{d\omega}{4\pi \imath T} \tanh \left( \frac{\omega}{2T}
\right) G_{1}^{+}(-\imath \omega )G_{2}^{+}(-\imath \omega -\imath
\Omega ).
\end{eqnarray}
Now we consider the derivative of the sum of $(S_1 + S_2 + S_3)$ with respect to Bosonic frequency
$\left. \frac{\partial }{\partial \Omega }\right\vert _{\Omega =0}$. For brevity we
omit arguments $-\imath \omega $ of Green's functions
\begin{eqnarray}
\label{Ssum}
 \left. \frac{\partial }{\partial \Omega }\right\vert _{\Omega =0}&&\left(
S_{1} + S_{2} + S_{3} \right) = \int \frac{d\omega}{4\pi \imath T} \tanh \left(
\frac{\omega}{2T}\right) \left[ \left( \frac{\partial }{\partial \omega }
G_{1}^{-}\right) G_{2}^{-} - G_{1}^{-}\left( \frac{\partial }{\partial \omega }
G_{2}^{+}\right) - \left( \frac{\partial }{\partial \omega }G_{1}^{-}\right)
G_{2}^{+} + G_{1}^{+}\left( \frac{\partial }{\partial \omega }G_{2}^{+}\right)
\right] \nonumber \\
&=& \int \frac{d\omega}{4\pi \imath T} \tanh \left( \frac{\omega}{2T}
\right)
\left[ \left( \frac{\partial }{\partial \omega }G_{1}^{-}\right) \left(
G_{2}^{-}-G_{2}^{+}\right) +\underset{\text{exchange indices}}{\underbrace{%
\left( \frac{\partial }{\partial \omega }G_{2}^{+}\right) \left(
G_{1}^{+}-G_{1}^{-}\right) }}\right] \nonumber \\
&=& \int \frac{d\omega}{4\pi \imath T} \tanh \left( \frac{\omega}{2T} \right)
\underset{\frac{1}{2}\left[ \left( G_{2}^{-}-G_{2}^{+}\right) \frac{\partial
}{\partial \omega }\left( G_{1}^{-}-G_{1}^{+}\right) +\left(
G_{1}^{-}-G_{1}^{+}\right) \frac{\partial }{\partial \omega }\left(
G_{2}^{-}-G_{2}^{+}\right) \right] }{\underbrace{\left(
G_{2}^{-}-G_{2}^{+}\right) \frac{\partial }{\partial \omega }\left(
G_{1}^{-}-G_{1}^{+}\right) }} \nonumber \\
&=& \int \frac{d\omega}{4\pi \imath T} \tanh \left( \frac{\omega}{2T}
\right) \frac{\partial }{\partial \omega }\left[ \left( G_{1}^{-}-G_{1}^{+}\right)
\left( G_{2}^{-}-G_{2}^{+}\right) \right]
= \int \frac{-d\omega}{4\pi \imath T} \tanh \left( \frac{\omega}{2T} \right)
\frac{\partial }{\partial \omega }\left[ \frac{1}{\tau _{\omega }^{2}}
G_{1}^{-}G_{1}^{+}G_{2}^{-}G_{2}^{+}\right].
\end{eqnarray}
Now one can perform the integration over variables $\xi_1$ and $\xi_2$
using Eqs.~(\ref{sum}), (\ref{Ssum}), and the residuum theorem
\begin{eqnarray}
&& \int d\xi _{1}d\xi _{2}\ g(\xi _{1},\xi
_{2})G_{1}^{-}G_{1}^{+}G_{2}^{-}G_{2}^{+} \\
&=&\int d\xi _{1}d\xi _{2}\ \frac{\xi _{1}+\xi _{2}+\frac{d}{2}\frac{%
\left( \xi _{1}+\xi _{2}\right) ^{2}}{\varepsilon _{F}}}{\left( \omega -\xi
_{1}-\imath /(2\tau )\right) \underset{\xi _{1,0}=\omega +\imath /(2\tau )}{%
\underbrace{\left( \omega -\xi _{1}+\imath /(2\tau )\right) }}\left( \omega
-\xi _{2}-\imath /(2\tau )\right) \left( \omega -\xi _{2}+\imath /(2\tau
)\right) } \nonumber \\
&=&2\pi \imath \int d\xi _{2}\frac{\omega +\imath /(2\tau )+\xi _{2}+\frac{d%
}{2\varepsilon _{F}}\left( \omega +\imath /(2\tau )+\xi _{2}\right) ^{2}}{%
\left( -\imath /\tau \right) \left( \omega -\xi _{2}-\imath /(2\tau )\right)
\underset{\xi _{2,0}=\omega +\imath /(2\tau )}{\underbrace{\left( \omega
-\xi _{2}+\imath /(2\tau )\right) }}}
= 4\pi ^{2}\tau ^{2}\left[ 2\omega +\imath /\tau +\frac{d}{2\varepsilon _{F}%
}\left( 2\omega +\imath /\tau \right) ^{2}\right]\,, \nonumber
\end{eqnarray}
where the $\xi_{i,0}$-values below the braces denote the poles in the complex plane used to perform the integration.
As a result we obtain the following expression for the derivative of correlation function
\begin{eqnarray}
\label{Q00}
\left. \frac{\partial }{\partial \Omega }\right\vert _{\Omega =0}Q^{(0)}
&=& \frac{s}{2d}et_{0}^{2} a^{2d+2}\left( \nu _{d}^{(0)}\right) ^{2}
\int \frac{d\omega}{8\pi \imath} \tanh \left( \frac{\omega}{2T}\right) \frac{%
\partial }{\partial \omega }\left[ 4\pi ^{2}\left[ 2\omega +\imath /\tau +%
\frac{d}{2\varepsilon _{F}}\left( 2\omega +\imath /\tau \right) ^{2}\right] %
\right] \\
&=&\frac{\pi s}{4\imath d}et_{0}^{2}a^{2d+2}\left( \nu _{d}^{(0)}\right)
^{2}\int d\omega \tanh \left( \frac{\omega}{2T}\right) \frac{\partial }{\partial
\omega }\left[ \frac{d}{2\varepsilon _{F}}\left( 2\omega +\imath /\tau _{0}%
\left[ (d/2-1)\omega /\varepsilon _{F}\right] \right) ^{2}\right] \nonumber \\
&=&-\frac{\pi s}{4\imath d}et_{0}^{2}a^{2d+2}\left( \nu _{d}^{(0)}\right)
^{2}\left( 2T\right) ^{-1}\frac{d}{2\varepsilon _{F}}\int d\omega \frac{%
\left( 2+(d/2-1)\imath /\left( \tau _{0}\varepsilon _{F}\right) \right)
^{2}\omega ^{2}}{\cosh ^{2}\left( \omega /(2T)\right) } \nonumber \\
&= &-\frac{\pi s}{4\imath d}et_{0}^{2}a^{2d+2}\left( \nu
_{d}^{(0)}\right) ^{2}\left( 2T\right) ^{2}\frac{4d}{2\varepsilon _{F}}\frac{
\pi ^{2}}{6}. \nonumber
\end{eqnarray}
\end{widetext}
In the second line in Eq.~(\ref{Q00}) the derivative is taken into account (removing the
boundary terms of the partial integration) and the contributions of order $
1/\varepsilon^2_{F}$ or smaller are neglected in the last line.

Finally, we obtain the following expression for non-interacting thermoelectric coefficient of granular metals
\begin{equation}
\label{eta0}
\eta ^{(0)}=-\frac{s\pi ^{3}}{3}et_{0}^{2}a^{d+2}\left( \nu
_{d}^{(0)}\right) ^{2}\frac{T}{\varepsilon _{F}}.
\end{equation}
One can re-write this expression using the relations: $\nu _{d}^{(0)}\mathcal{D}%
_{d}=ga^{2-d}$ ; \ $\nu _{d}^{(0)}=(\delta a^{d})^{-1}$ ; $t_{0}^{2}=g\delta
^{2}/(2\pi )$, where $\mathcal{D}_{d}$ is the diffusion constant, $g$ the
tunneling conductance, and $\delta $ the mean level spacing, giving
\begin{equation}\label{eq.eta0alt}
\eta^{(0)}= - \frac{s\pi^2}{6} eg_T a^{2-d} (T/\varepsilon _{F}).
\end{equation}

\section{Thermoelectric coefficient of granular metals with interaction}
\label{app.eta1}

In this appendix we consider the correction $\eta^{(1)}$ to the thermoelectric coefficient of granular metals due to electron-electron interaction, i.e.,
\begin{equation}
\eta=\eta^{(0)}+\eta^{(1)},
\end{equation}
where $\eta^{(0)}$ was calculated in Appendix~\ref{app.eta0}, Eq.~(\ref{eta0}).
The structure of the diagrams $Q^{(1)}$, $Q^{(2)}$, $Q^{(3)}$ contributing to $\eta^{(1)}$ are shown in Fig.~\ref{fig.dia} and we can write
\begin{equation}
\label{eta1sum}
\eta ^{(1)}=\imath \left. \frac{\partial }{a^{d}T\partial \Omega }%
\right\vert _{\Omega =0}\left( Q^{(1)}+Q^{(2)}+Q^{(3)}\right).
\end{equation}
These diagrams include the effect of elastic scattering of electron at impurities described by diffusons, $\mathcal{D}^{-1} = \tau _{\omega }\left( \left\vert \Omega _{i}\right\vert + \epsilon _{q}\delta \right)$, and the effect of the dynamically screened Coulomb potential $\widetilde{V}(q,\Omega _{i}) = \mathcal{D}V(q,\Omega _{i}) \mathcal{D}$:
\begin{eqnarray}
\widetilde{V}(q,\Omega _{i}) &=& \frac{2E_{c}(q)}{\tau _{\omega }^{2}\left[ \left\vert \Omega
_{i}\right\vert +4E_{c}(q)\epsilon _{q}\right] \left[ \left\vert \Omega
_{i}\right\vert +\epsilon _{q}\delta \right] }, \\
V(q,\Omega _{i}) &=&\left( \frac{1}{2E_{c}(q)}+\frac{2\epsilon _{q}}{
\left\vert \Omega _{i}\right\vert +\epsilon _{q}\delta }\right) ^{-1}. \nonumber
\end{eqnarray}
The renormalized interaction vertices are: (i) inter-grain
\begin{eqnarray}
\label{Phi1}
\Phi _{\omega }^{(1)}(\Omega _{i}) &=& a^{d}\int \frac{d%
\overrightarrow{q}}{(2\pi )^{d}}\widetilde{V}(q,\Omega _{i}) \sum_{a}^{\prime }\cos (%
\overrightarrow{q}\cdot \overrightarrow{a})   \\
&=& \int \frac{ a^{d} d\overrightarrow{%
q}}{(2\pi )^{d}} \frac{2 \tau _{\omega }^{-2}E_{c}(q)\sum_{a}^{\prime }\cos (\overrightarrow{q}%
\cdot \overrightarrow{a})}{\left[ \left\vert \Omega
_{i}\right\vert +4E_{c}(q)\epsilon _{q}\right] \left[ \left\vert \Omega
_{i}\right\vert +\epsilon _{q}\delta \right] }, \nonumber
\end{eqnarray}
and (ii) intra-grain:
\begin{eqnarray}
\label{Phi2}
\Phi _{\omega }^{(2)}(\Omega _{i}) &=& a^{d}\, 2d \int \frac{d%
\overrightarrow{q}}{(2\pi )^{d}}\widetilde{V}(q,\Omega _{i}) \\
&=& \int \frac{d^d d\overrightarrow{q}}{(2\pi )^{d}}\frac{4d \, \tau _{\omega }^{-2} E_{c}(q)}{\left[
\left\vert \Omega _{i}\right\vert +4E_{c}(q)\epsilon _{q}\right] \left[
\left\vert \Omega _{i}\right\vert +\epsilon _{q}\delta \right] }, \nonumber
\end{eqnarray}
with $\epsilon _{q}=2g_{T}\left( 2d-\sum_{a}^{\prime }\cos (%
\overrightarrow{q}\cdot \overrightarrow{a})\right) $ where $%
\sum_{a}^{\prime }$ stands for summation over all directions and
orientations $\left\{ \pm a\overrightarrow{e}_{j}^{(0)}\right\} $, and
\begin{equation}\label{eq.Ecq}
E_{c}(q)=\frac{e^{2}}{2C(q)}=\frac{e^{2}}{a^{d}}\left\{
\begin{array}{l}
-\ln (qa), \hspace{0.4cm} d=1 \\
\pi /q,  \hspace{1.1cm} d=2 \\
2\pi /q^{2}, \hspace{0.8cm} d=3.
\end{array}%
\right.
\end{equation}
\begin{figure}[t]
\includegraphics[width=0.8\linewidth]{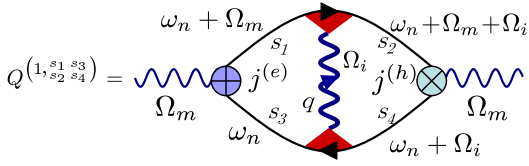}
\caption{(color online) Diagram describing the correction to the thermoelectric coefficient due to electron-electron interaction corresponding to the term $Q^{(1)}$ in Eq.~(\ref{eta1sum}). The external Bosonic frequency is denoted by $\Omega$ (wavy lines) and the internal Fermionic frequency by $\omega$ (straight lines).
The electric and heat current vertexes are $\protect\overrightarrow{\jmath_e}$ and $\protect\overrightarrow{\jmath_h}$ (without Coulomb contribution), respectively. The (red) triangles denote the diffusons $\mathcal{D}$ and the thick wavy line the screen Coulomb interaction.}\label{fig.Q1}
\end{figure}
Explicitly, the contribution $Q^{(1)}$ in Eq.~(\ref{eta1sum}) is given by
\begin{eqnarray}
\label{Q1def}
Q^{(1)} &=&-\frac{s}{2d}et_{0}^{2}T^{2}a^{2d+2}\left( \nu _{d}^{(0)}\right)
^{2} \\
&\times& \underset{\omega _{n},\Omega _{i}}{\sum }\int d\xi _{1}d\xi
_{2}g_{12} F_{1}^{\left( s_{1}s_{2}s_{3}s_{4}\right)
}\Phi _{\omega }^{(1)}\left( \Omega _{i}\right) \nonumber \\
&=& Q^{\left( 1,{\pm \pm }\right) }+Q^{\left( 1,{\mp \mp }\right) }\,, \nonumber
\end{eqnarray}
with $g_{12}\equiv g(\xi_1,\xi_2)$ defined in Eq.~(\ref{g12}) of Appendix~\ref{app.eta0}.
Here we introduce the function
\begin{eqnarray}
F_{1}^{\left( {s_{1}{s_{2}}}{s_{3}{s_{4}}}
\right) } &=& G_{1}^{s_{1}}(\omega _{n} + \Omega _{m})G_{1}^{s_{2}}(\omega
_{n}+\Omega _{m}+\Omega _{i}) \nonumber \\
&\times& G_{2}^{s_{3}}(\omega _{n})G_{2}^{s4}(\omega
_{n}+\Omega _{i}),
\end{eqnarray}
with $s_{i}$ denote the analytic structure of the Green's functions
implying restrictions on the frequency summation -- in principle there are 16 different combinations of the $s_i$'s, see Fig.~\ref{fig.Q1}. However, only the two diagrams $Q^{\left( 1,{\pm \pm }\right) }$ and $%
Q^{\left( 1,{\mp \mp }\right) }$ contribute to the correction $\eta^{(1)}$. The
other analytical structures do not have either valid frequency domains or
poles are located in only one half plane of $\mathbb{C}$.
\begin{figure}[t]
\includegraphics[width=0.8\linewidth]{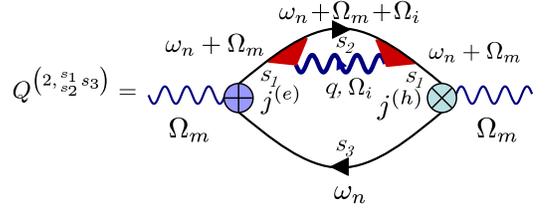}
\caption{(color online) Diagram describing correction to the thermoelectric coefficient due to electron-electron interaction corresponding to term $Q^{(2)}$ in Eq.~(\ref{eta1sum}). All notations are the same as in Fig.~\ref{fig.Q1}}\label{fig.Q2}
\end{figure}
For the contribution $Q^{(2)}$ in Eq.~(\ref{eta1sum}), Fig.~\ref{fig.Q2}, we have the following expression
\begin{eqnarray}
\label{Q2}
&& Q^{(2)} = -\frac{s}{2d}et_{0}^{2}T^{2}a^{2d+2}\left( \nu _{d}^{(0)}\right)
^{2}\underset{\omega _{n},\Omega _{i}}{\sum }\int d\xi _{1}d\xi
_{2}g_{12} \\
&& \times F_{2}^{\left( {{s_{1}}{s_{2}}}s_{3}\right) }\Phi _{\omega}^{(2)}\left( \Omega _{i}\right)
= 2\left[ Q^{\left( 2,{\pm +}\right) }+Q^{\left( 2,{\mp -}\right)}+Q^{\left( 2,{\pm -}\right) }\right],
\nonumber
\end{eqnarray}
where
\begin{equation}
F_{2}^{\left( {{s_{1}}{s_{2}}}s_{3}\right) }=\left[
G_{1}^{s_{1}}(\omega _{n}+\Omega _{m})\right] ^{2}G_{1}^{s_{2}}(\omega
_{n}+\Omega _{m}+\Omega _{i})G_{2}^{s_{3}}(\omega _{n}).
\end{equation}
Due to symmetry all three contributing diagrams in the right hand side of Eq.~(\ref{Q2}) have
a factor of 2. Again, out of the eight possible combinations for $s_1$, $s_2$, and $s_3$ only three combinations
 -- $Q^{\left( 2,{\pm +}\right) }$, $Q^{\left( 2,{\pm -}\right) }$, $Q^{\left( 2,{\mp -}\right) }$ -- have a valid or non-zero analytical structure.
\begin{figure}[t]
\includegraphics[width=0.8\linewidth]{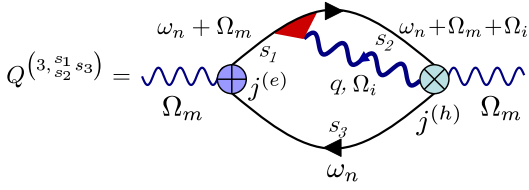}
\caption{(color online) Diagram describing correction to the thermoelectric coefficient due to electron-electron interaction corresponding to term $Q^{(3)}$ in Eq.~(\ref{eta1sum}). All notations are the same as in Fig.~\ref{fig.Q1}, but here the heat vertex corresponds to $\widehat{\jmath}^{h,1}_{i}$ and the diagram has only one diffuson.}\label{fig.Q3}
\end{figure}

The diagram $Q^{(3)}$, shown in Fig.~\ref{fig.Q3}, describes the contribution of the correlation function with the interaction part of the heat current operator, $\widehat{\jmath }_{ij}^{(h,1)}$ (see Appendix~\ref{app.cur}), and has therefore a different structure in comparison with contributions $Q^{(1)}$ and $Q^{(2)}$:
\begin{eqnarray}
Q^{(3)} &=&-\frac{s}{2d}et_{0}^{2}T^{2}a^{2d+2}\left( \nu _{d}^{(0)}\right)
^{2} \\
&& \times \underset{\omega _{n},\Omega _{i}}{\sum }\int d\xi _{1}d\xi
_{2}g_{3} F_{3}^{\left( {s_{1}{s_{2}}}s_{3}\right) }\Phi _{3}\left(\Omega _{i},q\right), \nonumber
\end{eqnarray}
with
\begin{equation}
F_{3}^{\left( {s_{1}{s_{2}}}s_{3}\right) }=G_{1}^{s_{1}}(\omega _{n}+\Omega _{m}+\Omega _{i})G_{1}^{s_{2}}(\omega
_{n}+\Omega _{m})G_{2}^{s_{3}}(\omega _{n}),
\end{equation}
and
\begin{eqnarray}
&& g_{3}\left( \xi _{1},\xi_{2}\right) =2\left( 1+\frac{\xi _{1}+\xi _{2}}{\varepsilon _{F}}\right)
\\
&& \times \left( 1+\left( \frac{d}{2}-1\right) \frac{\xi _{1}}{\varepsilon _{F}}
\right) \left( 1+\left( \frac{d}{2}-1\right) \frac{\xi _{2}}{\varepsilon _{F}
}\right) \nonumber \\
&& = 2\left( 1+\frac{d}{2\varepsilon _{F}}\left( \xi _{1}+\xi
_{2}\right) \right) + O(\xi ^{2}/\varepsilon _{F}^{2}). \nonumber
\end{eqnarray}
Since the linear part of function $g_3(\xi _{1},\xi_{2})$ has a factor $\varepsilon _{F}^{-1}$, the main contribution to $\eta^{(1)}$ from the diagram $Q^{(3)}$ is of the order of $T^{2}/\varepsilon _{F}^{2}$,
whereas  $Q^{(1)}$ and  $Q^{(2)}$ have $1/\varepsilon _{F}$ contributions, which we are considering here only.
Therefore we will not consider diagram $Q^{(3)}$ any further.

In the following we discuss the five diagrams contributing to $\eta^{(1)}$ in details, especially their analytical structure and the resulting restrictions on the frequency summations:
$Q^{\left( 1,{\pm \pm }\right) }$, $Q^{\left( 1,{\mp \mp }\right) }$, $Q^{\left( 2,{\pm +}\right) }$, $Q^{\left( 2,{\pm -}\right) }$, and $Q^{\left( 2,{\mp -}\right) }$.

\subsection{Calculation of contribution $Q^{\left( 1,{\pm \pm }\right) }$ in Eq.~(\ref{Q1def}).}\label{app.eta1.s1}

Here we discuss the contribution $Q^{\left( 1,{\pm \pm }\right) }$ introduced in Eq.~(\ref{Q1def}).
The analytical structure of this diagram (Fig.~\ref{fig.Q1}), defined by indexes $s_1$ to $s_4$
in Eq.~(\ref{Q1def}), demands
\begin{eqnarray}
\omega _{n}+\Omega _{m} >0\quad&,&\qquad \omega _{n}+\Omega _{m}+\Omega _{i} <0 \\
\omega _{n} >0\quad&,&\qquad \omega _{n}+\Omega _{i} <0\,.\nn
\end{eqnarray}
These inequalities define the limits of frequency summations:
\begin{equation}
0<\omega _{n}<-\Omega_{m}-\Omega _{i}\,\text{ and }\, \Omega _{i}<-\Omega _{m}\,.
\end{equation}
First, we calculate the $\xi $-integrals in Eq.~(\ref{Q1def}) using the residue theorem:
\begin{widetext}
\begin{eqnarray}
&&\int d\xi _{1}d\xi _{2}\frac{\xi _{1}+\xi _{2}+\frac{d}{2}\frac{\left( \xi
_{1}+\xi _{2}\right) ^{2}}{\varepsilon _{F}}}{\left( \imath (\omega
_{n}+\Omega _{m})-\xi _{1}+\imath /(2\tau )\right) \underset{\xi
_{1,0}=\imath (\omega _{n}+\Omega _{m}+\Omega _{i})-\imath /(2\tau )\in
\mathbb{C}^{-}}{\underbrace{\left( \imath (\omega _{n}+\Omega _{m}+\Omega _{i})-\xi
_{1}-\imath /(2\tau )\right) }}\underset{\xi _{2,0}=\imath \omega
_{n}+\imath /(2\tau )\in \mathbb{C}^{+}}{\underbrace{\left( \imath \omega _{n}-\xi _{2}+\imath /(2\tau )\right)
}}\left( \imath (\omega _{n}+\Omega _{i})-\xi _{2}-\imath /(2\tau )\right) }
\nonumber \\
&=&-(2\pi \imath )^2 \frac{\imath \left( 2\omega _{n}+\Omega
_{m}+\Omega _{i}\right) -\frac{d}{2\varepsilon _{F}}\left( 2\omega
_{n}+\Omega _{m}+\Omega _{i}\right) ^{2}}{\left( -\imath \Omega _{i}+\imath
/\tau \right) \left( \imath \Omega _{i}-\imath /\tau \right) } =
\frac{4\pi ^{2}}{\left( \Omega _{i}-1/\tau \right) ^{2}}\left[ \imath
\left( 2\omega _{n}+\Omega _{m}+\Omega _{i}\right) -\frac{d}{2\varepsilon
_{F}}\left( 2\omega _{n}+\Omega _{m}+\Omega _{i}\right) ^{2}\right] \nonumber \\
&=&\frac{4\pi ^{2}}{\left( \Omega _{i}-1/\tau \right) ^{2}}g\left( \imath
\left( 2\omega _{n}+\Omega _{m}+\Omega _{i}\right) \right) \approx 4\pi
^{2}\tau ^{2}g\left( \imath \left( 2\omega _{n}+\Omega _{m}+\Omega
_{i}\right) \right),
\label{xi1xi2}
\end{eqnarray}
\end{widetext}
where in the last line we introduce the function
\begin{equation}
g(z)=z + [d/(2\varepsilon _{F})] z^{2}.
\end{equation}
For the final approximation we
used the fact that $\left\vert \Omega _{i}\right\vert \tau \ll 1$. The poles for
the residual are written below the underbraces.

Using the result of integration over $\xi_1$ and $\xi_2$ in Eq.~(\ref{xi1xi2}), we can simplify the expression for this diagram to
\begin{eqnarray}
\label{11}
Q^{\left( 1,{\pm \pm }\right) } = -\frac{2 \pi^2 s}{d}et_{0}^{2}T^{2}\tau^2 a^{2d+2}
\left( \nu_{d}^{(0)} \right)^2 \\
\underset{ \underset{ \Omega_{i} < - \Omega_m }{0 < \omega_n < - \Omega_m - \Omega_ i} }
{\times\sum } \Phi_{\omega }^{(1)}\left( \Omega_{i}\right) g\left( \imath \left( 2 \omega_{n} + \Omega_{m} + \Omega_{i} \right) \right)\,.\nn
\end{eqnarray}

\subsection{Calculation of contribution $Q^{\left( 1,{\mp \mp }\right) }$ in Eq.~(\ref{Q1def})}
\label{app.eta1.s2}

Here we discuss the contribution $Q^{\left( 1,{\mp \mp }\right) }$ defind in Eq.~(\ref{Q1def}).
The analytical structure of this diagram demands
\begin{eqnarray}
\omega _{n}+\Omega _{m} <0 \quad&,&\qquad
\omega _{n}+\Omega _{m}+\Omega _{i} >0 \\
\omega _{n} <0 \quad&,&\qquad\omega _{n}+\Omega _{i} >0\,,\nn
\end{eqnarray}
which defines the limits of the frequency summations:
\begin{equation}
-\Omega _{i}<\omega_{n}<-\Omega _{m}\,\text{ and }\,\Omega _{i}>\Omega _{m}\,.
\end{equation}
We first perform the $\xi $-integrals in Eq.~(\ref{Q1def}):
\begin{widetext}
\begin{eqnarray}
\label{Q1--}
\int \frac{d\xi _{1}d\xi _{2} \left(\xi _{1}+\xi _{2}+\frac{d}{2}\frac{\left( \xi
_{1}+\xi _{2}\right) ^{2}}{\varepsilon _{F}}\right) }{\left[ \imath (\omega
_{n}+\Omega _{m})-\xi _{1}-\imath /(2\tau )\right] \underset{\xi
_{1,0}=\imath (\omega _{n}+\Omega _{m}+\Omega _{i})+\imath /(2\tau )\in
\mathbb{C}^{+}}{\underbrace{\left[ \imath (\omega _{n}+\Omega _{m}+\Omega _{i})-\xi
_{1}+\imath /(2\tau )\right] }}\underset{\xi _{2,0}=\imath \omega
_{n}-\imath /(2\tau )\in
\mathbb{C}^{-}}{\underbrace{\left[ \imath \omega _{n}-\xi _{2}-\imath /(2\tau )\right]
}}\left[ \imath (\omega _{n}+\Omega _{i})-\xi _{2}+\imath /(2\tau )\right] }
\nonumber \\
= 4\pi^2 \frac{\imath \left( 2\omega _{n}+\Omega
_{m}+\Omega _{i}\right) -\frac{d}{2\varepsilon _{F}}\left( 2\omega
_{n}+\Omega _{m}+\Omega _{i}\right) ^{2}}{\left( -\imath \Omega _{i}-\imath
/\tau \right) \left( \imath \Omega _{i}+\imath /\tau \right) } = \frac{4\pi ^{2} g\left( \imath
\left( 2\omega _{n}+\Omega _{m}+\Omega _{i}\right) \right)}{\left( \Omega _{i}+1/\tau \right) ^{2}}
 \approx 4\pi
^{2}\tau ^{2}g\left( \imath \left( 2\omega _{n}+\Omega _{m}+\Omega
_{i}\right) \right).
\end{eqnarray}
\end{widetext}
Substituting the result of Eq.~(\ref{Q1--}) back into Eq.~(\ref{Q1def}) we obtain
\begin{eqnarray}
\label{12}
Q^{\left( 1,{\mp \mp }\right) } = -\frac{\pi^2 s}{d}et_{0}^{2}T^{2}\tau^2 a^{2d+2}
\left( \nu _{d}^{(0)}\right)^{2}  \\
\times \underset{\underset{\Omega _{i}>\Omega_{m}}
{-\Omega _{i}<\omega _jvv{n} < - \Omega_{m}}}{\sum } \Phi_{\omega }^{(1)}
\left( \Omega _{i}\right) g\left( \imath \left( 2\omega_{n}+\Omega _{m} + \Omega_{i}\right) \right). \nonumber
\end{eqnarray}

\bigskip

\subsection{Calculation of contribution $Q^{\left( 2,{\pm +}\right) }$ in Eq.~(\ref{Q2})}
\label{app.eta1.s3}

Here we discuss the contribution $Q^{\left( 2,{\pm + }\right) }$ introduced in Eq.~(\ref{Q2}).
The analytical structure of this diagram demands
\begin{eqnarray}
\omega _{n}+\Omega _{m} &>&0\,,\,\,\omega _{n}+\Omega _{m}+\Omega _{i} <0\nn \\
\omega _{n} &>&0
\end{eqnarray}
which defines the limits of frequency summations:
\begin{equation}
0<\omega _{n}<-\Omega_{m}-\Omega _{i}\,\text{ and }\,\Omega _{i}<-\Omega _{m}.
\end{equation}
For symmetry reasons we write both versions of the integrals (diffusons on
grain 1 \& 2) and therefore already take the factor 2 in front of the $
Q^{(2)}$\ subdiagrams into account in the right hand side of Eq.~(\ref{Q2}). Furthermore we introduce the short
notation for the Green's functions:
\begin{eqnarray}
\label{notations}
a_{\pm }(\xi ) &\equiv &\imath (\omega _{n}+\Omega _{m})-\xi \pm \imath
/(2\tau ), \\
b_{\pm }(\xi ) &\equiv &\imath (\omega _{n}+\Omega _{m}+\Omega _{i})-\xi \pm
\imath /(2\tau ), \nonumber \\
c_{\pm }(\xi ) &\equiv &\imath \omega _{n}-\xi \pm \imath /(2\tau ). \nonumber
\end{eqnarray}
In the following we need only the poles of functions $b_{\pm }(\xi )$ and $c_{\pm }(\xi )$: $\xi _{\pm
}^{(b)}=\imath (\omega _{n}+\Omega _{m}+\Omega _{i})\pm \imath /(2\tau )\in
\mathbb{C}^{\pm }$ and $\xi _{\pm }^{(c)}=\imath \omega _{n}\pm \imath /(2\tau )\in\mathbb{C}^{\pm }$. We also use the
function $g_{12}$ introduced in Eq.~(\ref{Q2}).

Therefore the $\xi $-integrals in Eq.~(\ref{Q2}) can be written as:
\begin{widetext}
\begin{eqnarray}
&&\int d\xi _{1}d\xi _{2}g\left( \xi _{1}+\xi _{2}\right) \left[ \left(
a_{+}^{2}(\xi _{1})b_{-}(\xi _{1})c_{+}(\xi _{2})\right) ^{-1}+\left(
a_{+}^{2}(\xi _{2})b_{-}(\xi _{2})c_{+}(\xi _{1})\right) ^{-1}\right] \\
&=&\int d\xi _{1}d\xi _{2}g\left( \xi _{1}+\xi _{2}\right) \frac{%
a_{+}^{2}(\xi _{2})b_{-}(\xi _{2})c_{+}(\xi _{1})+a_{+}^{2}(\xi
_{1})b_{-}(\xi _{1})c_{+}(\xi _{2})}{a_{+}^{2}(\xi _{1})b_{-}(\xi
_{1})c_{+}(\xi _{1})a_{+}^{2}(\xi _{2})b_{-}(\xi _{2})c_{+}(\xi _{2})} \nonumber \\
&=& 2\pi \imath \int d\xi _{1}\frac{1}{a_{+}^{2}(\xi _{1})b_{-}
(\xi_{1})c_{+}(\xi _{1})}\left[ g\left( \xi _{1}+\xi _{+}^{\left( c\right)
}\right) \frac{c_{+}(\xi _{1})}{2}-\frac{g\left( \xi _{1}+\xi _{-}^{\left(
b\right) }\right) a_{+}^{2}(\xi _{1})b_{-}(\xi _{1})}{a_{+}^{2}\left( \xi
_{-}^{\left( b\right) }\right) }\right] \nonumber \\
&=& 2\pi \imath \int d\xi _{1}\left[ \frac{g\left( \xi _{1}+\xi
_{+}^{\left( c\right) }\right) }{2a_{+}^{2}(\xi _{1})b_{-}(\xi _{1})}-\frac{
g\left( \xi _{1}+\xi _{-}^{\left( b\right) }\right) }{a_{+}^{2}\left( \xi
_{-}^{\left( b\right) }\right) c_{+}(\xi _{1})}\right]. \nonumber
\end{eqnarray}
\end{widetext}
Here we executed the $\xi _{2}$-integral and used the fact that the first term in the
second line has only one pole in $\mathbb{C}^{+}$ (factor $\pi \imath $). We are left with two term where the second one
has only a single $\xi _{1}$-pole in $\mathbb{C}^{+}$ which gives another factor $\pi \imath $. However, both terms give the
same result
\begin{eqnarray}
\label{integration}
&& 4 \pi^2 \frac{g\left( \xi _{+}^{\left( c\right)
} + \xi _{-}^{\left( b\right) }\right) }{a_{+}^{2}\left( \xi _{-}^{\left(
b\right) }\right) } = \frac{4\pi ^{2} g\left( \imath \left( 2\omega _{n}+\Omega _{m}+\Omega
_{i}\right) \right)}{\left( -\imath \Omega _{i}+\imath /\tau
\right) ^{2}} \nonumber \\
&\approx &-4\pi ^{2}\tau ^{2}g\left( \imath \left( 2\omega _{n}+\Omega
_{m}+\Omega _{i}\right) \right).
\end{eqnarray}
Substituting the result of Eq.~(\ref{integration}) into  Eq.~(\ref{Q2}) we obtain
\begin{eqnarray}
\label{13}
2Q^{\left( 2,{\pm +}\right) }=  \frac{2\pi^2 s}{2d}et_{0}^{2}T^{2}\tau^2 a^{2d+2}
\left( \nu _{d}^{(0)}\right) ^{2} \\
\times \underset{\underset{\Omega _{i}<-\Omega _{m}
}{0<\omega _{n}<-\Omega _{m}-\Omega _{i}}}{\sum } \Phi
_{\omega }^{(2)}\left( \Omega _{i}\right) g\left( \imath \left( 2\omega
_{n}+\Omega _{m}+\Omega_{i}\right) \right). \nonumber
\end{eqnarray}
The notation introduced in this Appendix allows us to write down the $Q^{\left( 2,{\mp -}\right) }$ and
$Q^{\left( 2,{\pm -}\right) }$ contributions in Eq.~(\ref{Q2}) by just changing the +/- indices.

\bigskip

\subsection{Calculation of contribution $Q^{\left( 2,{\mp -}\right) }$ in Eq.~(\ref{Q2})}
\label{app.eta1.s4}

Here we discuss the contribution $Q^{\left( 2,{\mp - }\right) }$ introduced in Eq.~(\ref{Q2}).
The analytical structure of this diagram demands
\begin{eqnarray}
\omega _{n}+\Omega _{m} &<&0\,,\,\,\omega _{n}+\Omega _{m}+\Omega_{i} >0, \\
\omega _{n} &<&0, \nonumber
\end{eqnarray}
which defines the limits of frequency summations:
\begin{equation}
-\Omega _{m}-\Omega_{i}<\omega_{n}<-\Omega_{m}\,\text{ and }\,\Omega_{i}>0.
\end{equation}
Using notations introduced in Eq.~(\ref{notations}) the $\xi $-integrals in Eq.~(\ref{Q2}) can be calculated as:
\begin{widetext}
\begin{eqnarray}
\label{integration2}
&&\int d\xi _{1}d\xi _{2}g\left( \xi _{1}+\xi _{2}\right) \left[ \left(
a_{-}^{2}(\xi _{1})b_{+}(\xi _{1})c_{-}(\xi _{2})\right) ^{-1}+\left(
a_{-}^{2}(\xi _{2})b_{+}(\xi _{2})c_{-}(\xi _{1})\right) ^{-1}\right] \\
&=&\int d\xi _{1}d\xi _{2}g\left( \xi _{1}+\xi _{2}\right) \frac{%
a_{-}^{2}(\xi _{2})b_{+}(\xi _{2})c_{-}(\xi _{1})+a_{-}^{2}(\xi
_{1})b_{+}(\xi _{1})c_{-}(\xi _{2})}{a_{-}^{2}(\xi _{1})b_{+}(\xi
_{1})c_{-}(\xi _{1})a_{-}^{2}(\xi _{2})b_{+}(\xi _{2})c_{-}(\xi _{2})} \nonumber \\
&=& 2\pi \imath \int d\xi _{1}\frac{1}{a_{-}^{2}(\xi _{1})b_{+}(\xi
_{1})c_{-}(\xi _{1})}\left[ -g\left( \xi _{1}+\xi _{-}^{\left( c\right)
}\right) \frac{c_{-}(\xi _{1})}{2}+\frac{g\left( \xi _{1}+\xi _{+}^{\left(
b\right) }\right) a_{-}^{2}(\xi _{1})b_{+}(\xi _{1})}{a_{-}^{2}\left( \xi
_{+}^{\left( b\right) }\right) }\right] \nonumber \\
&=& 2\pi \imath \int d\xi _{1}\left[ -\frac{g\left( \xi _{1}+\xi
_{-}^{\left( c\right) }\right) }{2a_{-}^{2}(\xi _{1})b_{+}(\xi _{1})}+\frac{%
g\left( \xi _{1}+\xi _{+}^{\left( b\right) }\right) }{a_{-}^{2}\left( \xi
_{+}^{\left( b\right) }\right) c_{-}(\xi _{1})}\right]
= 4\pi^2 \frac{g\left( \xi _{-}^{\left( c\right)
}+\xi _{+}^{\left( b\right) }\right) }{a_{-}^{2}\left( \xi _{+}^{\left(
b\right) }\right) }\approx -4\pi ^{2}\tau ^{2}g\left( \imath \left( 2\omega
_{n}+\Omega _{m}+\Omega _{i}\right) \right). \nonumber
\end{eqnarray}
\end{widetext}
Substituting the result of Eq.~(\ref{integration2}) into  Eq.~(\ref{Q2}) we obtain
\begin{eqnarray}
\label{14}
2Q^{\left( 2,{\mp -}\right) } = \frac{2\pi^2 s}{d}et_{0}^{2}T^{2}\tau^2 a^{2d+2}
\left( \nu _{d}^{(0)}\right)^{2} \\
\times \underset{\underset{\Omega _{i}>0}{-\Omega
_{m}-\Omega_{i}<\omega _{n}<-\Omega_{m}}}{\sum } \Phi
_{\omega }^{(2)}\left( \Omega_{i}\right) g \left( \imath \left( 2\omega
_{n}+\Omega_{m}+\Omega_{i}\right) \right). \nonumber
\end{eqnarray}

\bigskip

\subsection{Calculation of contribution $Q^{\left( 2,{\pm -}\right) }$ in Eq.~(\ref{Q2})}
\label{app.eta1.s5}

Here we discuss the contribution $Q^{\left( 2,{\pm - }\right) }$ introduced in Eq.~(\ref{Q2}).
The analytical structure of this diagram demands
\begin{eqnarray}
\omega _{n}+\Omega _{m} &>&0 \,,\,\,\omega _{n}+\Omega _{m}+\Omega _{i} <0, \\
\omega _{n} &<&0,\nn
\end{eqnarray}
which defines the limits of frequency summations:
\begin{equation}
-\Omega _{m}<\omega
_{n}<0 \,\text{ and }\,\Omega _{i}<-\Omega _{m},
\end{equation}
and the disjunct region
\begin{equation}
-\Omega_{m}<\omega _{n}<-\Omega _{m}-\Omega _{i}\,\text{ and }\,-\Omega _{m}<\Omega _{i}<0.
\end{equation}
Using notations introduced in Eq.~(\ref{notations}) the $\xi $-integrals in Eq.~(\ref{Q2}) for this diagram
can be calculated as:
\begin{widetext}
\begin{eqnarray}
\label{integration3}
&&\int d\xi _{1}d\xi _{2}g\left( \xi _{1}+\xi _{2}\right) \left[ \left(
a_{+}^{2}(\xi _{1})b_{-}(\xi _{1})c_{-}(\xi _{2})\right) ^{-1}+\left(
a_{+}^{2}(\xi _{2})b_{-}(\xi _{2})c_{-}(\xi _{1})\right) ^{-1}\right] \\
&=&\int d\xi _{1}d\xi _{2}g\left( \xi _{1}+\xi _{2}\right) \frac{%
a_{+}^{2}(\xi _{2})b_{-}(\xi _{2})c_{-}(\xi _{1})+a_{+}^{2}(\xi
_{1})b_{-}(\xi _{1})c_{-}(\xi _{2})}{a_{+}^{2}(\xi _{1})b_{-}(\xi
_{1})c_{-}(\xi _{1})a_{+}^{2}(\xi _{2})b_{-}(\xi _{2})c_{-}(\xi _{2})} \nonumber \\
&=& 2\pi \imath \int d\xi _{1}\frac{1}{a_{+}^{2}(\xi _{1})b_{-}(\xi
_{1})c_{-}(\xi _{1})}\left[ -g\left( \xi _{1}+\xi _{-}^{\left( c\right)
}\right) \frac{c_{-}(\xi _{1})}{2}+\frac{g\left( \xi _{1}+\xi _{-}^{\left(
b\right) }\right) a_{+}^{2}(\xi _{1})b_{-}(\xi _{1})}{a_{+}^{2}\left( \xi
_{-}^{\left( b\right) }\right) }\right] \nonumber \\
&=& 2\pi \imath \int d\xi _{1}\left[ -\frac{g\left( \xi _{1}+\xi
_{-}^{\left( c\right) }\right) }{2a_{+}^{2}(\xi _{1})b_{-}(\xi _{1})}-\frac{%
g\left( \xi _{1}+\xi _{-}^{\left( b\right) }\right) }{a_{+}^{2}\left( \xi
_{-}^{\left( b\right) }\right) c_{-}(\xi _{1})}\right]
= -4 \pi^2 \frac{g\left( \xi _{-}^{\left( c\right) }+\xi
_{-}^{\left( b\right) }\right) }{a_{+}^{2}\left( \xi _{-}^{\left( b\right)
}\right) }\approx 4\pi ^{2}\tau ^{2}g\left( \imath \left( 2\omega
_{n}+\Omega _{m}+\Omega _{i}-1/\tau \right) \right). \nonumber
\end{eqnarray}
Substituting the result of Eq.~(\ref{integration3}) into  Eq.~(\ref{Q2}) taking into account the two disjunct regions for the frequency summations, we obtain
\begin{equation}
\label{15}
2Q^{\left( 2,{\pm -}\right) } =  \frac{2\pi^2 s}{d}et_{0}^{2}T^{2}\tau^2 a^{2d+2}
\left( \nu _{d}^{(0)}\right) ^{2}\left( \underset{\underset{\Omega
_{i}<-\Omega _{m}}{-\Omega _{m}<\omega _{n}<0}}{\sum }+\underset{\underset{
-\Omega _{m}<\Omega _{i}<0}{-\Omega _{m}<\omega _{n}<-\Omega _{m}-\Omega _{i}
}}{\sum }\right) \Phi _{\omega }^{(2)}\left( \Omega
_{i}\right) g\left( \imath \left( 2\omega _{n}+\Omega _{m}+\Omega
_{i}-1/\tau \right) \right).
\end{equation}

\subsection{Analytical continuation}

Here we combine all five contributions to calculate correction $\eta^{(1)}$: $Q^{\left( 1,{\pm \pm }\right)},
Q^{\left( 1,{\mp \mp }\right)}, Q^{\left( 2,{\pm +}\right)}, Q^{\left( 2,{\mp -}\right)}$, and
$Q^{\left( 2,{\pm -}\right)}$ introduced in the right hand sides of Eqs.~(\ref{Q1def}), (\ref{Q2}) and discussed in the previous subsections~(\ref{app.eta1.s1}) through~(\ref{app.eta1.s5}). We focus in particular on the analytical continuation of the Matsubara to real frequencies, both for the Fermionic and Bosonic frequencies.
Using Eqs.~(\ref{11}), (\ref{12}), (\ref{13}), (\ref{14}), and (\ref{15}) one can write
\begin{eqnarray}
\label{Qsum}
&& Q^{(1)} + Q^{(2)} = Q^{\left( 1,{\pm \pm }\right) }+Q^{\left( 1,{\mp \mp }\right) }+2\left(
Q^{\left( 2,{\pm +}\right) }+Q^{\left( 2,{\mp -}\right) }+Q^{\left( 2,{\pm -}%
\right) }\right) \\
&=&\lambda T\left[ -\underset{\underset{\Omega _{i}<-\Omega _{m}}{0<\omega
_{n}<-\Omega _{m}-\Omega _{i}}}{\sum }\tau ^{2}\Phi _{\omega }^{(1)}\left(
\Omega _{i}\right) g\left( \imath \left( 2\omega _{n}+\Omega_{m}+\Omega
_{i}\right) \right) -\underset{\underset{\Omega _{i}>\Omega_{m}}{-\Omega
_{i}<\omega_{n}<-\Omega_{m}}}{\sum }\tau ^{2}\Phi_{\omega }^{(1)}\left(
\Omega_{i}\right) g\left( \imath \left( 2\omega _{n}+\Omega_{m}+\Omega
_{i}\right) \right) \right. \nonumber \\
&+&\underset{\underset{\Omega_{i}<-\Omega _{m}}{0<\omega _{n}<-\Omega
_{m}-\Omega _{i}}}{\sum }\tau ^{2}\Phi _{\omega }^{(2)}\left( \Omega
_{i}\right) g\left( \imath \left( 2\omega _{n}+\Omega _{m}+\Omega
_{i}\right) \right) +\underset{\underset{\Omega _{i}>0}{-\Omega _{m}-\Omega
_{i}<\omega _{n}<-\Omega _{m}}}{\sum }\tau ^{2}\Phi _{\omega }^{(2)}\left(
\Omega _{i}\right) g\left( \imath \left( 2\omega _{n}+\Omega _{m}+\Omega
_{i}\right) \right) \nonumber \\
&-&\left. \underset{\underset{\Omega _{i}<-\Omega _{m}}{-\Omega _{m}<\omega
_{n}<0}}{\sum }\tau ^{2}\Phi _{\omega }^{(2)}\left( \Omega _{i}\right)
g\left( \imath \left( 2\omega _{n}+\Omega _{m}+\Omega _{i}-1/\tau \right)
\right) -\underset{\underset{-\Omega _{m}<\Omega _{i}<0}{-\Omega _{m}<\omega
_{n}<-\Omega _{m}-\Omega _{i}}}{\sum }\tau ^{2}\Phi _{\omega }^{(2)}\left(
\Omega _{i}\right) g\left( \imath \left( 2\omega _{n}+\Omega _{m}+\Omega
_{i}-1/\tau \right) \right) \right], \nonumber
\end{eqnarray}
where we introduced the notation $\lambda =\frac{2\pi ^{2}s}{d}et_{0}^{2}Ta^{2d+2}\left( \nu
_{d}^{(0)}\right)^{2}$ and the functions $\Phi _{\omega }^{(1)}\left(
\Omega _{i}\right)$ and $\Phi _{\omega }^{(2)}\left(\Omega _{i}\right)$ were defined
in Eqs.~(\ref{Phi1}) and (\ref{Phi2}) respectively.

Next, we perform the summation over Fermionic frequencies $\omega _{n}$ in Eq.~(\ref{Qsum}) by shifts and the corresponding analytical continuation
\begin{eqnarray}
\underset{\underset{\Omega _{i}<-\Omega _{m}}{0<\omega _{n}<-\Omega
_{m}-\Omega _{i}}}{\sum }f(\omega _{n},\Omega _{i}) &=&\left[ \underset{%
\underset{\Omega _{i}<-\Omega _{m}}{\omega _{n}>0}}{\sum }-\underset{%
\underset{\Omega _{i}<-\Omega _{m}}{\omega _{n}>-\Omega _{m}-\Omega _{i}}}{%
\sum }\right] f(\omega _{n},\Omega _{i})=\underset{\underset{\Omega
_{i}<-\Omega _{m}}{\omega _{n}>0}}{\sum }\left[ f(\omega _{n},\Omega
_{i})-f(\omega _{n}-\Omega _{m}-\Omega _{i},\Omega _{i})\right], \\
\underset{\underset{\Omega _{i}>\Omega _{m}}{-\Omega _{i}<\omega
_{n}<-\Omega _{m}}}{\sum }f(\omega _{n},\Omega _{i}) &=&\left[ \underset{%
\underset{\Omega _{i}>\Omega _{m}}{\omega _{n}<-\Omega _{m}}}{\sum }-%
\underset{\underset{\Omega _{i}>\Omega _{m}}{\omega _{n}<-\Omega i}}{\sum }%
\right] f(\omega _{n},\Omega _{i})=\underset{\underset{\Omega _{i}>\Omega
_{m}}{\omega _{n}<0}}{\sum }\left[ f(\omega _{n}-\Omega _{m},\Omega
_{i})-f(\omega _{n}-\Omega _{i},\Omega _{i})\right], \nonumber \\
\underset{\underset{\Omega _{i}>0}{-\Omega _{m}-\Omega _{i}<\omega
_{n}<-\Omega _{m}}}{\sum }f(\omega _{n},\Omega _{i}) &=&\left[ \underset{%
\underset{\Omega _{i}>0}{\omega _{n}<-\Omega _{m}}}{\sum }-\underset{%
\underset{\Omega _{i}>0}{\omega _{n}<-\Omega _{m}-\Omega _{i}}}{\sum }\right]
f(\omega _{n},\Omega _{i})=\underset{\underset{\Omega _{i}>0}{\omega _{n}<0}}%
{\sum }\left[ f(\omega _{n}-\Omega _{m},\Omega _{i})-f(\omega _{n}-\Omega
_{i}-\Omega _{m},\Omega _{i})\right], \nonumber \\
\underset{\underset{\Omega _{i}<-\Omega _{m}}{-\Omega _{m}<\omega _{n}<0}}{%
\sum }f(\omega _{n},\Omega _{i}) &=&\left[ \underset{\underset{\Omega
_{i}<-\Omega _{m}}{\omega _{n}<0}}{\sum }-\underset{\underset{\Omega
_{i}<-\Omega _{m}}{\omega _{n}<-\Omega _{m}}}{\sum }\right] f(\omega
_{n},\Omega _{i})=\underset{\underset{\Omega _{i}<-\Omega _{m}}{\omega _{n}<0%
}}{\sum }\left[ f(\omega _{n},\Omega _{i})-f(\omega _{n}-\Omega _{m},\Omega
_{i})\right], \nonumber \\
\underset{\underset{-\Omega _{m}<\Omega _{i}<0}{-\Omega _{m}<\omega
_{n}<-\Omega _{m}-\Omega _{i}}}{\sum }f(\omega _{n},\Omega _{i}) &=&\left[
\underset{\underset{-\Omega _{m}<\Omega _{i}<0}{\omega _{n}<-\Omega
_{m}-\Omega _{i}}}{\sum }-\underset{\underset{-\Omega _{m}<\Omega _{i}<0}{%
\omega _{n}<-\Omega _{m}}}{\sum }\right] f(\omega _{n},\Omega _{i}) \nonumber \\
&=& \underset{\underset{-\Omega _{m}<\Omega _{i}<0}{\omega _{n}<0}}{\sum }\left[ f(\omega
_{n}-\Omega _{m}-\Omega _{i},\Omega _{i})-f(\omega _{n}-\Omega _{m},\Omega
_{i})\right] \nonumber \\
&=&\underset{\underset{\Omega _{i}<0}{\omega _{n}<0}}{\sum }\left[ f(\omega
_{n}-\Omega _{m}-\Omega _{i},\Omega _{i})-f(\omega _{n}-\Omega _{m},\Omega
_{i})\right] \nonumber \\
&-& \underset{\underset{\Omega _{i}<-\Omega _{m}}{\omega _{n}<0}}{%
\sum }\left[ f(\omega _{n}-\Omega _{m}-\Omega _{i},\Omega _{i})-f(\omega
_{n}-\Omega _{m},\Omega _{i})\right]. \nonumber
\end{eqnarray}
\end{widetext}
Here the function $f(\omega _{n},\Omega _{i})$ is the product of the functions $\tau^2 \Phi _{\omega }^{(\alpha)}\left(
\Omega _{i}\right)$ and $g$. For the analytic continuation we need to consider the $\omega$
-dependence of function $\Phi _{\omega }^{(\alpha)}\left(\Omega _{i}\right)$, in particular we can use the fact that (which follows directly from the definition of functions $\Phi _{0}^{(\alpha )}$ in Eqs.~(\ref{Phi1}) and (\ref{Phi2}))
\begin{equation}
\label{d40}
\Phi _{0}^{(\alpha )}\tau _{0}^{2}=\Phi _{\omega }^{(\alpha )}\tau _{\omega
}^{2},
\end{equation}
where $\Phi _{0}^{(\alpha )}$ is the same as $\Phi _{\omega }^{(\alpha )}$
but with $\tau _{\omega }$ replaced by $\tau _{0}$ and is therefore $\omega
$-independent. Using Eq.~(\ref{d40}), we obtain
\begin{widetext}
\begin{eqnarray}
\label{111}
&&Q^{(1)} + Q^{(2)}  = Q^{\left( 1,{\pm \pm }\right) }+Q^{\left( 1,{\mp \mp }\right) }+2\left(
Q^{\left( 2,{\pm +}\right) }+Q^{\left( 2,{\mp -}\right) }+Q^{\left( 2,{\pm -}%
\right) }\right) \\
= \frac{\lambda T}{4\pi \imath T} &&\left[ -\underset{\Omega _{i}<-\Omega _{m}%
}{\sum }\left[ \Phi _{0}^{(1)}\left( \Omega _{i}\right) -\Phi
_{0}^{(2)}\left( \Omega _{i}\right) \right] \tau _{0}^{2}\int d\omega \tanh
\left(\frac{\omega}{2T}\right)\left[ g\left( 2\omega +\imath \left( \Omega _{m}+\Omega
_{i}\right) \right) -g\left( 2\omega -\imath \left( \Omega _{m}+\Omega
_{i}\right) \right) \right] \right. \nonumber \\
&& + \underset{\Omega _{i}>\Omega _{m}}{\sum }\Phi _{0}^{(1)}\left( \Omega
_{i}\right) \tau _{0}^{2}\int d\omega \tanh \left(\frac{\omega}{2T}\right) \left[ g\left(
2\omega +\imath \left( -\Omega _{m}+\Omega _{i}\right) \right) -g\left(
2\omega +\imath \left( \Omega _{m}-\Omega _{i}\right) \right) \right] \nonumber \\
&&-\underset{\Omega _{i}>0}{\sum }\Phi _{0}^{(2)}\left( \Omega _{i}\right)
\tau _{0}^{2}\int d\omega \tanh \left(\frac{\omega}{2T}\right) \left[ g\left( 2\omega +\imath
\left( -\Omega _{m}+\Omega _{i}\right) \right) -g\left( 2\omega -\imath
\left( \Omega _{m}+\Omega _{i}\right) \right) \right] \nonumber \\
&&+\underset{\Omega _{i}<-\Omega _{m}}{\sum }\Phi _{0}^{(2)}\left( \Omega
_{i}\right) \tau _{0}^{2}\int d\omega \tanh \left(\frac{\omega}{2T}\right) \left[ g\left(
2\omega +\imath \left( \Omega _{m}+\Omega _{i}-1/\tau \right) \right)
-g\left( 2\omega +\imath \left( -\Omega _{m}+\Omega _{i}-1/\tau \right)
\right) \right] \nonumber \\
&&+\underset{\Omega _{i}<0}{\sum }\Phi _{0}^{(2)}\left( \Omega _{i}\right)
\tau _{0}^{2}\int d\omega \tanh \left(\frac{\omega}{2T}\right) \left[ g\left( 2\omega -\imath
\left( \Omega _{m}+\Omega _{i}+1/\tau \right) \right) -g\left( 2\omega
+\imath \left( -\Omega _{m}+\Omega _{i}-1/\tau \right) \right) \right] \nonumber \\
&&\left. -\underset{\Omega _{i}<-\Omega _{m}}{\sum }\Phi _{0}^{(2)}\left(
\Omega _{i}\right) \tau _{0}^{2}\int d\omega \tanh
\left(\frac{\omega}{2T}\right) \left[ g\left( 2\omega -\imath \left( \Omega _{m}+\Omega _{i}+1/\tau \right)
\right) -g\left( 2\omega +\imath \left( -\Omega _{m}+\Omega _{i}-1/\tau
\right) \right) \right] \right]. \nonumber
\end{eqnarray}
Next, we consider the integrands in Eq.~(\ref{111}) and introduce the short hand notations
\begin{eqnarray}
\label{112}
a_{1}\equiv g\left( 2\omega +\imath \left( \Omega _{m}+\Omega _{i}\right)
\right) -g\left( 2\omega -\imath \left( \Omega _{m}+\Omega _{i}\right)
\right) &=&2\imath \left( \Omega _{m}+\Omega _{i}\right) \left( 1+2d\omega
/\varepsilon _{F}\right), \\
a_{2}\equiv g\left( 2\omega +\imath \left( -\Omega _{m}+\Omega _{i}\right)
\right) -g\left( 2\omega +\imath \left( \Omega _{m}-\Omega _{i}\right)
\right) &=&2\imath \left( \Omega _{i}-\Omega _{m}\right) \left( 1+2d\omega
/\varepsilon _{F}\right), \nonumber \\
a_{3}\equiv g\left( 2\omega +\imath \left( -\Omega _{m}+\Omega _{i}\right)
\right) -g\left( 2\omega -\imath \left( \Omega _{m}+\Omega _{i}\right)
\right) &=&2\imath \Omega _{i}+2\Omega _{i}d\left( 2\omega \imath +\Omega
_{m}\right) /\varepsilon _{F}, \nonumber \\
a_{4}\equiv g\left( 2\omega +\imath \left( \Omega _{m}+\Omega _{i}-1/\tau
\right) \right) -g\left( 2\omega +\imath \left( -\Omega _{m}+\Omega
_{i}-1/\tau \right) \right) &=&2\Omega _{m}\imath +2\Omega _{m}d/\varepsilon
_{F}\left( 1/\tau _{\omega } + 2\imath \omega -\Omega _{i}\right), \nonumber \\
a_{5}\equiv g\left( 2\omega -\imath \left( \Omega _{m}+\Omega _{i}+1/\tau
\right) \right) -g\left( 2\omega +\imath \left( -\Omega _{m}+\Omega
_{i}-1/\tau \right) \right) &=&-2\imath \Omega_{i}-2d\Omega
_{i}/\varepsilon_{F}\left( 1/\tau _{\omega }+2\imath \omega -\Omega
_{m}\right).  \nonumber
\end{eqnarray}
\end{widetext}
Now, we extract only terms which are linear in $\omega $ in Eq.~(\ref{112}) and of order $%
1/\varepsilon _{F}$ (thus the $\tau _{\omega }$ in $a_{4}$ and $a_{5}$\ does
not give a contribution). Therefore we obtain
\begin{eqnarray}
\label{114}
a_{1} &\simeq &2\omega \imath \left( \Omega _{m}+\Omega _{i}\right)
2d/\varepsilon _{F} \\
a_{2} &\simeq &2\omega \imath \left( \Omega _{i}-\Omega _{m}\right)
2d/\varepsilon _{F} \nonumber \\
a_{3} &\simeq &2\omega \imath \Omega _{i}2d/\varepsilon _{F} \nonumber \\
a_{4} &\simeq &2\omega \imath \Omega _{m}2d/\varepsilon _{F} \nonumber \\
a_{5} &\simeq &-2\omega \imath \Omega _{i}2d/\varepsilon _{F}. \nonumber
\end{eqnarray}
Substituting the result of Eqs.~(\ref{114}) back into Eq.~(\ref{111}) we obtain
\begin{widetext}
\begin{eqnarray}
\label{116}
&&Q^{(1)} + Q^{(2)} = Q^{\left( 1,{\pm \pm }\right) }+Q^{\left( 1,{\mp \mp }\right) }+2\left(
Q^{\left( 2,{\pm +}\right) }+Q^{\left( 2,{\mp -}\right) }+Q^{\left( 2,{\pm -}%
\right) }\right) \\
=\frac{2\tau _{0}^{2}\imath 2d\lambda T}{4\pi \imath T\varepsilon _{F}} &&%
\left[ -\underset{\Omega _{i}<-\Omega _{m}}{\sum }\left[ \Phi
_{0}^{(1)}\left( \Omega _{i}\right) -\Phi _{0}^{(2)}\left( \Omega
_{i}\right) \right] \int d\omega \tanh (\omega /2T)\omega \left( \Omega
_{m}+\Omega _{i}\right) \right. \nonumber \\
&&+\underset{\Omega _{i}>\Omega _{m}}{\sum }\Phi _{0}^{(1)}\left( \Omega
_{i}\right) \int d\omega \tanh (\omega /2T)\omega \left( \Omega _{i}-\Omega
_{m}\right) -\underset{\Omega _{i}>0}{\sum }\Phi _{\omega }^{(2)}\left( \Omega
_{i}\right) \int d\omega \tanh (\omega /2T)\omega \Omega _{i} \nonumber \\
&&+\underset{\Omega _{i}<-\Omega _{m}}{\sum }\Phi _{0}^{(2)}\left( \Omega
_{i}\right) \int d\omega \tanh (\omega /2T)\omega \Omega _{m} -
\underset{\Omega _{i}<0}{\sum }\Phi _{0}^{(2)}\left( \Omega _{i}\right)
\int d\omega \tanh (\omega /2T)\omega \Omega _{i} \nonumber \\
&&\left. +\underset{\Omega _{i}<-\Omega _{m}}{\sum }\Phi _{0}^{(2)}\left(
\Omega _{i}\right) \int d\omega \tanh (\omega /2T)\omega \Omega _{i}\right]
\nonumber \\
=\widetilde{\lambda } &&\left[ -\underset{\Omega _{i}<-\Omega _{m}}{\sum }%
\left[ \Phi _{0}^{(1)}\left( \Omega _{i}\right) -\Phi _{0}^{(2)}\left(
\Omega _{i}\right) \right] \left( \Omega _{m}+\Omega _{i}\right)
\right. +\underset{\Omega _{i}>\Omega _{m}}{\sum }\Phi _{0}^{(1)}\left( \Omega
_{i}\right) \left( \Omega _{i}-\Omega _{m}\right) -\underset{\Omega _{i}>0}{%
\sum }\Phi _{0}^{(2)}\left( \Omega _{i}\right) \Omega _{i} \nonumber \\
&&\left. +\underset{\Omega _{i}<-\Omega _{m}}{\sum }\Phi _{0}^{(2)}\left(
\Omega _{i}\right) \Omega _{m}-\underset{\Omega _{i}<0}{\sum }\Phi
_{0}^{(2)}\left( \Omega _{i}\right) \Omega _{i}+\underset{\Omega
_{i}<-\Omega _{m}}{\sum }\Phi _{0}^{(2)}\left( \Omega _{i}\right) \Omega _{i}%
\right]. \nonumber
\end{eqnarray}
\end{widetext}
In the last two lines of Eq.~(\ref{116}) we introduced the notation
$\widetilde{\lambda }=\frac{\tau _{0}^{2}d\lambda }{\pi \varepsilon _{F}%
}\int d\omega \omega \tanh (\omega /2T)$ where we use the fact that
$\int d\omega \omega \tanh (\omega /2T)\longrightarrow =-\frac{(\pi
T)^{2}}{3}$, neglecting the infinite boundary terms of the partial
integration.

Using the fact that the functions $\Phi_{0}^{(2)}\left(\Omega _{i}\right)$
in Eq.~(\ref{116}) depend only on the absolute value of the Bosonic frequency $\Omega$,
we finally obtain
\begin{eqnarray}
\label{Q1Q2}
&& Q^{\left( 1\right) }+Q^{\left( 2\right) } \\
&& = -2 \lambda_1 \underset{\Omega _{i}<-\Omega _{m}}{\sum }\left[ \Phi _{0}^{(1)}\left(
\Omega _{i}\right) -\Phi _{0}^{(2)}\left( \Omega _{i}\right) \right] \left(
\Omega _{m}+\Omega _{i}\right), \nonumber
\end{eqnarray}
where we use the notation $\lambda_1 =-\frac{2\pi ^{3}s}{3}e\left( \tau _{0}t_{0}\nu
_{d}^{(0)}a^{d+1}\right) ^{2}\frac{T^{3}}{\varepsilon _{F}}$.
The internal frequency summation is also done by analytical continuation,
but for Bosonic frequencies: $\Omega _{i}\longrightarrow -\imath \widetilde{%
\Omega }+\eta $. Before the final integration we should take the external
frequency derivative and finally calculate the ${\bf q}$-integrals of the $\Phi_0 $%
-functions. Using Eqs.~(\ref{Phi1}) and (\ref{Phi2}) we have
\begin{eqnarray}
\label{difference}
\Phi _{0}^{(1)}\left( \Omega _{i}\right) -\Phi _{0}^{(2)}\left( \Omega
_{i}\right) = a^{d}\int \frac{d {\bf q}}{(2\pi )^{d}} \\
\times \frac{2E_{c}(q)
\left[ 2d-\sum_{a}^{^{\prime }}\cos ({\bf q}\cdot {\bf
a})\right] }{\tau_{0}^{2}\left( \left\vert \Omega _{i}\right\vert
+4E_{c}(q)\epsilon_{\bf q}\right) \left( \left\vert \Omega _{i}\right\vert
+\epsilon _{\bf q}\delta \right) }. \nonumber
\end{eqnarray}
Using Eq.~(\ref{difference}) one can calculate the sum over internal bosonic frequencies
$\Omega_i$ in Eq.~(\ref{Q1Q2})
\begin{eqnarray}
\label{S}
S_{\Omega } &=&\underset{\Omega _{i}<-\Omega _{m}}{\sum }\frac{\Omega
_{m}+\Omega _{i}}{\left( \left\vert \Omega _{i}\right\vert
+4E_{c}(q)\epsilon _{q}\right) \left( \left\vert \Omega _{i}\right\vert
+\epsilon _{q}\delta \right) } \\
&=&\underset{\Omega _{i}<-\Omega _{m}}{\sum }\frac{\Omega _{m}+\Omega _{i}}{%
\left( -\Omega _{i}+4E_{c}(q)\epsilon _{q}\right) \left( -\Omega
_{i}+\epsilon _{q}\delta \right) } \nonumber \\
&=&\underset{\Omega _{i}<0}{\sum }\frac{\Omega _{i}}{\left( -\Omega
_{i}+\Omega _{m}+4E_{c}(q)\epsilon _{q}\right) \left( -\Omega _{i}+\Omega
_{m}+\epsilon _{q}\delta \right) } \nonumber \\
&=& \int \frac{ [-4\pi \imath T]^{-1} ( -\imath \widetilde{\Omega })
\coth (\widetilde{\Omega }/2T)d\widetilde{\Omega }}{[
\imath \widetilde{\Omega } +\Omega _{m}+4E_{c}(q)\epsilon _{q} ]
[ \imath \widetilde{\Omega } +\Omega _{m}+\epsilon _{q}\delta] } \nonumber \\
&=& \int \frac{[-4\pi T]^{-1} \widetilde{\Omega }\coth (\widetilde{\Omega }%
/2T)d\widetilde{\Omega }}{[ \widetilde{\Omega }-\imath \left( \Omega
_{m}+4E_{c}(q)\epsilon _{q}\right)] [ \widetilde{\Omega }-\imath
\left( \Omega _{m}+\epsilon _{q}\delta \right)] } \nonumber \\
&=& \int \frac{ [-4\pi T]^{-1} x\coth (x)dx}{[ x-\frac{\imath }{2T}%
\left( \Omega _{m}+4E_{c}(q)\epsilon _{q}\right)] [ x-\frac{%
\imath }{2T}\left( \Omega _{m}+\epsilon _{q}\delta \right)] }. \nonumber
\end{eqnarray}
Only the terms proportional to the external frequency $\Omega_{m}$
in Eq.~(\ref{S}) contribute to the correction to the thermoelectric
coefficient $\eta^{(1)}$ in Eq.~(\ref{eta1sum}). Taking the derivative of both
sides of Eq.~(\ref{S}) we obtain
\begin{widetext}
\begin{eqnarray}
\label{Sderivative}
\left. \frac{\partial }{\partial \Omega }\right\vert _{\Omega =0}S_{\Omega }
&=& -\frac{1}{4\pi T}\int \left. \frac{\partial }{\partial \Omega }%
\right\vert _{\Omega =0}\frac{x\coth (x)dx}{\left( x-\frac{\Omega }{2T}-%
\frac{\imath }{T}2E_{c}(q)\epsilon _{q}\right) \left( x-\frac{\Omega }{2T}-%
\frac{\imath }{2T}\epsilon _{q}\delta \right) } \\
&=& -\frac{1}{8\pi T^{2}}\int x\coth (x)dx\left[ \frac{1}{\left( x-\imath
a\right) ^{2}\left( x-\imath b\right) }+\frac{1}{\left( x-\imath a\right)
\left( x-\imath b\right) ^{2}}\right] \nonumber \\
&=& -\frac{1}{8\pi T^{2}}\int \frac{x\coth (x)dx\left[ 2x-\imath \left(
a+b\right) \right] }{\left( x-\imath a\right) ^{2}\left( x-\imath b\right)
^{2}}\equiv -\frac{1}{8\pi T^{2}}I_{a,b}, \nonumber
\end{eqnarray}
\end{widetext}
where $a=2E_{c}(q)\epsilon _{q}/T$\ and $b=\epsilon _{q}\delta /(2T)$.
Finally, we use the approximation:
\begin{equation}
x\coth (x)\approx \left\{
\begin{array}{ll}
1, & \left\vert x\right\vert <1 \\
\left\vert x\right\vert , & \left\vert x\right\vert \geq 1,
\end{array}%
\right.
\end{equation}
and obtain for the integral $I_{a,b}$ in Eq.~(\ref{Sderivative}) the following result
\begin{equation}
I_{a,b}= \imath \frac{\ln \left[ \left( 1+a^{2}\right) /\left(
1+b^{2}\right) \right] }{\left( a-b\right) }.
\end{equation}
For $a\gg 1\gg b$
\begin{equation}
I_{a,b} = \imath \frac{2\ln a}{a}=\imath \frac{T\ln \left[
2E_{c}(q)\epsilon _{q}/T\right] }{E_{c}(q)\epsilon _{q}}.
\end{equation}
Thus, for Eq.~(\ref{Sderivative}) we obtain
\begin{eqnarray}
\label{S2}
\left. \frac{\partial }{\partial \Omega }\right\vert _{\Omega =0}S_{\Omega }
= \frac{-\imath }{8\pi T}\frac{\ln \left[ 2E_{c}(q)\epsilon _{q}/T
\right] }{E_{c}(q)\epsilon _{q}}.
\end{eqnarray}
Substituting Eq.~(\ref{S2}) into Eq.~(\ref{difference}) we obtain for the correction to the thermoelectric
coefficient the following result
\begin{eqnarray}
\label{d52}
\eta ^{(1)} &=& - \frac{\widetilde{\lambda }}{2\pi
T^{2}(2\pi )^{d}\tau _{0}^{2}} \\
&\times& \int d {\bf q} \frac{\left[
2d-\sum_{a}^{^{\prime }}\cos ({\bf q}\cdot {\bf a})
\right] \ln \left[ 2E_{c}(q)\epsilon_{\bf q}/T\right] }{\epsilon _{\bf q}}. \nonumber
\end{eqnarray}

\subsection{Final integration and results}

In Eq.~(\ref{d52}) we are left with the final integration over internal momenta ${\bf q}$.
Therefore we need the functional dependence of $\epsilon_{\bf q}$ and $E_{c}(q)$ on ${\bf q}$, which were introduced around Eq.~(\ref{eq.Ecq}) as
\begin{eqnarray}
\epsilon _{q} &=&2g_{T}\left( 2d-\sum_{a}^{\prime}\cos ({\bf q
}\cdot {\bf a})\right) \\
E_{c}(q) &=&\frac{e^{2}}{2C(q)}=\frac{e^{2}}{a^{d}}\left\{
\begin{array}{l}
-\ln (qa),d=1 \\
\pi /q,d=2 \\
2\pi /q^{2},d=3%
\end{array}%
\right. .
\end{eqnarray}
The final ${\bf q}$-integral can therefore be written as
\begin{equation}\label{eq.eta1.finalint}
\eta ^{(1)} = \frac{-\widetilde{\lambda }}{4\pi T^{2}(2\pi
)^{d}g_{T}\tau _{0}^{2}}\int d {\bf q}\ln \left[ 2E_{c}(q)\epsilon_{\bf q}/T\right]\,.
\end{equation}
The $q$-integral is cutoff at $q=\pi /a$. In $d=3$ the $\ln $
-argument is finite at $q=0$. In $d=2$ the volume element makes the
integrand at $q=0$ finite. The coefficient in Eq.~(\ref{eq.eta1.finalint}) can be simplified to
\begin{eqnarray}
\frac{-\widetilde{\lambda }}{4\pi T^{2}(2\pi )^{d}g_{T}\tau _{0}^{2}} &=&%
\frac{\pi ^{2}s}{6(2\pi )^{d}}\frac{et_{0}^{2}a^{2d+2}}{g_{T}}\left( \nu
_{d}^{(0)}\right) ^{2}\frac{T}{\varepsilon _{F}} \nonumber \\
&=& -\frac{\eta ^{(0)}}{2\pi g_{T}}\left( \frac{a}{2\pi }\right) ^{d}\,,
\end{eqnarray}
such that in $d=3$ we obtain:
\begin{eqnarray}
\eta _{3D}^{(1)} &=& -\frac{\eta ^{(0)}}{2\pi g_{T}}\left( \frac{a}{%
2\pi }\right)^{3} \int d{\bf q} \\
&\times& \ln \left[\frac{16\pi e^2 g_T}{a T}  \frac{3-\cos (aq_{x})-\cos (aq_{y})-\cos (aq_{z})}{(aq)^2} \right] \nonumber \\
&= &-\frac{\eta^{(0)}}{2\pi g_{T}}\left( \frac{a}{2\pi }\right) ^{3}\frac{
\pi^{3}}{a^{3}}\left[ \frac{4\pi }{3}\ln \left( 16\frac{e^{2}}{a}g_{T}\pi
/T\right) + c_{3}\right]. \nonumber
\end{eqnarray}
Here $c_{3}=\int d{\bf q}\ln \left[ \left( 3 - \cos (\pi
q_{x})-\cos (\pi q_{y})-\cos (\pi q_{z})\right) /\left( \pi q\right)^{2}
\right] $ and integration is over the unit sphere. However, this numerical constant (and all
other inside the logarithm) can be neglected, since $E_{c}/T\gg 1$ and $g_{T}\gg 1$.

For $d=2$ we obtain
\begin{eqnarray}
\eta _{2D}^{(1)} & =& \frac{-\eta ^{(0)}}{2\pi g_{T}}\left( \frac{a}{
2\pi }\right)^{2} \int d {\bf q} \\
&&\times \ln \left[ \frac{8e^{2}g_{T}\pi }{
Ta}\left( 3-\cos (aq_{x})-\cos (aq_{y})\right) /\left( aq\right) \right] \nonumber \\
&=& \frac{-\eta ^{(0)}}{2\pi g_{T}}\left( \frac{a}{2\pi }\right) ^{2}\frac{
\pi ^{3}}{a^{2}}\left[ \ln \frac{8e^{2}g_{T}\pi }{Ta}+c_{2}\right], \nonumber
\end{eqnarray}
with $c_{2}=\int \frac{d{\bf q}}{\pi}\ln \left[ \frac{2-\cos (\pi
q_{x})-\cos (\pi q_{y}}{\pi q} \right]$.

\subsubsection{Final results}

Now, we can summarize our final results in $d=2,3$ for the thermoelectric coefficient (only the functional dependencies of the correction terms are kept under the logarithms):
\begin{eqnarray}
\eta ^{(0)} &=&-\frac{s\pi ^{3}}{3}et_{0}^{2}a^{d+2}\left( \nu
_{d}^{(0)}\right) ^{2}\frac{T}{\varepsilon _{F}}, \\
\eta _{3D}^{(1)} &=& \frac{s\pi ^{3}}{54}\frac{ea^{5}}{g_{T}}
t_{0}^2\left( \nu _{3}^{(0)}\right) ^{2}\frac{T}{\varepsilon _{F}}\ln \left( \frac{
e^{2}g_{T}}{Ta}\right)  \\
&=& -\frac{\eta ^{(0)}}{12 g_{T}} \ln \left( E_{c}g_{T}/T\right), \nonumber \\
\eta _{2D}^{(1)} &=& \frac{s\pi ^{3}}{24}\frac{ea^{4}}{g_{T}}\left(
t_{0}\nu _{2}^{(0)}\right) ^{2}\frac{T}{\varepsilon _{F}}\ln \left( \frac{%
e^{2}g_{T}}{Ta}\right)  \\
&=& -\frac{\eta ^{(0)}}{8 g_{T}} \ln \left( E_{c}g_{T}/T\right)\,, \nonumber
\end{eqnarray}
or combined in a compact way (valid for $d=2,3$) we obtain the following
final result for the thermoelectric coefficient of granular metals
\begin{equation}
\eta =\eta ^{(0)}\left( 1 - \frac{1}{4\pi g_{T}}\ln \frac{g_{T} E_c}{T} \right)\,.
\end{equation}

\end{document}